\documentclass[twocolumn,trackchanges]{aastex701}

\usepackage{xspace}

\newcommand{\Teff}{\ensuremath{T_\mathrm{eff}}\xspace}
\newcommand{\logg}{\ensuremath{\log{g}}\xspace}

\newcommand{\vsini}{\ensuremath{v \sin i}\xspace}

\shorttitle{TIDES I: Eccentric and Misaligned Binaries}
\shortauthors{Morgan et al.}

\begin{document}

\title{Tracing the Influence of Distant companions on Exoplanetary Systems (TIDES) \\
 I. Highly Eccentric and Misaligned Binary Companions to Hot Jupiters}

\newcommand{\caltech}{Department of Astronomy, California Institute of Technology, Pasadena, CA 91125, USA}
\newcommand{\gps}{Division of Geological \& Planetary Sciences, California Institute of Technology, Pasadena, CA 91125, USA}
\newcommand{\ucsc}{Department of Astronomy \& Astrophysics, University of California, Santa Cruz, CA95064, USA}
\newcommand{\keck}{W. M. Keck Observatory, Kamuela, HI, USA}
\newcommand{\ucla}{Department of Physics \& Astronomy, University of California, Los Angeles, CA 90095, USA}
\newcommand{\uclagps}{Department of Earth, Planetary, and Space Sciences, University of California, Los Angeles, CA 90095, USA}
\newcommand{\jpl}{Jet Propulsion Laboratory, California Institute of Technology, 4800 Oak Grove Dr.,Pasadena, CA 91109, USA}
\newcommand{\ucsd}{Department of Astronomy \& Astrophysics,  University of California, San Diego, La Jolla, CA 92093, USA}
\newcommand{\ifahonolulu}{Institute for Astronomy, University of Hawai`i, 2680 Woodlawn Drive, Honolulu, HI 96822, USA}
\newcommand{\berkeley}{Department of Astronomy, University of California at Berkeley, CA 94720, USA}
\newcommand{\ames}{NASA Ames Research Center, MS 245-6, Moffett Field, CA 94035, USA}
\newcommand{\osu}{Department of Astronomy, The Ohio State University, 100 W 18th Ave, Columbus, OH 43210 USA}
\newcommand{\ifahilo}{Institute for Astronomy, University of Hawaii at Hilo, 640 N Aohoku Pl, Hilo, HI 96720, USA}
\newcommand{\nmsu}{Department of Astronomy, New Mexico State University, PO Box 30001, MSC 4500, Las Cruces, NM 88003, USA}
\newcommand{\bern}{Center for Space and Habitability, University of Bern, Gesellschaftsstrasse 6, 3012 Bern, Switzerland}
\newcommand{\mpia}{Max-Planck-Institut für Astronomie, Königstuhl 17, 69117 Heidelberg, Germany}
\newcommand{\carnegiew}{Earth and Planets Laboratory, Carnegie Institution for Science, Washington, DC, 20015}
\newcommand{\toronto}{David A. Dunlap Institute Department of Astronomy \& Astrophysics, University of Toronto, Toronto, ON M5S 3H4, Canada}
\newcommand{\USQ}{Centre for Astrophysics, University of Southern Queensland, Toowoomba, QLD, Australia}
\newcommand{\upenn}{Department of Physics and Astronomy, University of Pennsylvania, Philadelphia PA 19104}
\newcommand{\kansas}{Department of Physics and Astronomy, University of Kansas, Lawrence, KS 66045, USA}
\newcommand{\jhu}{Department of Physics \& Astronomy, Johns Hopkins University, Baltimore, MD 21218, USA}
\newcommand{\nexsci}{NASA Exoplanet Science Institute, IPAC, Caltech, Pasadena, CA 91125, USA}
\newcommand{\stsci}{Space Telescope Science Institute, Baltimore, MD 21218, USA}
\newcommand{\arizona}{Department of Astronomy/Steward Observatory, University of Arizona, Tucson, AZ 85721, USA}
\newcommand{\herzberg}{NRC Herzberg Astronomy and Astrophysics, Victoria, BC, V9E 2E7, Canada}
\newcommand{\victoria}{Department of Physics and Astronomy, University of Victoria, Victoria, BC, V8P 5C2, Canada}
\newcommand{\umich}{Department of Astronomy, University of Michigan, Ann Arbor, MI 48109, USA}
\newcommand{\ipac}{IPAC, California Institute of Technology, Pasadena, CA 91125, USA}
\newcommand{\ciera}{Center for Interdisciplinary Exploration and Research in Astrophysics, Northwestern University, 1800 Sherman Ave, Evanston, IL 60201, USA}
\newcommand{\northwestern}{Department of Physics and Astronomy, Northwestern University, 2145 Sheridan Road, Evanston, IL 60208-3112}
\newcommand{\ucsb}{Department of Physics, University of California, Santa Barbara, CA 93106, USA}

\newcommand{\todo}[1]{\textcolor{red}{[#1]}}

\author[0000-0003-4022-6234]{Marvin Morgan}
\affiliation{\ucsb}
\email[show]{marvinmorgan@ucsb.edu}  

\author[0000-0002-6618-1137]{Jerry W. Xuan}
\altaffiliation{51 Pegasi b Fellow}
\affiliation{\ucsb}
\affiliation{\uclagps}
\email{jxuan@ucsb.edu}

\author[0000-0003-2649-2288]{Brendan P. Bowler}
\affiliation{\ucsb}
\email{bpbowler@ucsb.edu}

\author[0000-0003-4557-414X]{Kyle Franson}
\altaffiliation{NHFP Sagan Fellow}
\affiliation{\ucsc}
\email{kfranson@ucsc.edu}

\author[0000-0003-2232-7664]{Michael C. Liu}
\affiliation{\ifahonolulu}
\email{mliu@ifa.hawaii.edu}

\author[0000-0002-7094-7908]{Konstantin Batygin}
\affiliation{\gps}
\email{kbatygin@caltech.edu}

\author[0000-0002-5375-4725]{Heather A. Knutson}
\affiliation{\gps}
\email{hknutso2@caltech.edu}

\author[0000-0001-5578-1498]{Björn Benneke}
\affiliation{\uclagps}
\email{bbenneke@epss.ucla.edu}

\author[0000-0003-1172-5755]{Julianne Cronin}
\affiliation{\northwestern}
\email{JulianneCronin2021@u.northwestern.edu}

\author[0000-0002-1392-0768]{Luke Finnerty}
\affiliation{\umich}
\email{lfinnert@umich.edu}

\author[0000-0002-5370-7494]{Chih-Chun Hsu}
\affiliation{\northwestern}
\email{chsu@northwestern.edu}

\author[0000-0002-1838-4757]{Aniket Sanghi}
\altaffiliation{NSF Graduate Research Fellow}
\affiliation{\caltech}
\email{asanghi@caltech.edu}

\begin{abstract}

The origin of hot Jupiters remains an open question. A significant fraction reside in binary or multiple star systems, suggesting that stellar companions may play an important role in shaping their present-day architectures. However, the orbits of the vast majority of these stellar companions remain poorly constrained, leaving their role in hot Jupiter evolution an open question. We present the Tracing the Influence of Distant companions on Exoplanetary Systems (TIDES) survey, a campaign to map the orbits of stellar companions in systems hosting  hot Jupiters to assess the extent to which binaries play a role in hot Jupiter formation and migration. Here, we introduce our survey design and new orbital fits for the first six systems in our sample—$\tau$ Boo, WASP-2, WASP-11, WASP-76, HAT-P-14, and HAT-P-16—by combining previous observations of the stellar companions together with new Keck/NIRC2 imaging and Keck/KPIC relative radial velocities. Among these six systems, the four with the most precisely constrained orbits exhibit high eccentricities ($e \geq 0.8$), and three of these systems also display significant planet--binary misalignments ($\Delta i \geq 20^\circ$). 
This suggests the orbital properties of stellar companions to hot Jupiter host stars are qualitatively different than field binaries and binaries hosting inner transiting small planets. These orbital architectures are consistent with stellar companions shaping the evolution of hot Jupiters, either through high-eccentricity migration or by participating in an earlier episode of dynamical evolution that ultimately led to hot Jupiter formation. Continued orbital monitoring of the full sample will reveal whether outer stellar companions play a causal role in the evolution of hot Jupiters.

\end{abstract}

\keywords{Exoplanet Formation -- Exoplanet migration -- Low mass stars -- Multiple stars}

\section{Introduction} 

While the frequency of planets around stars beyond the Sun has long been a fundamental question in astronomy (\citealt{Struve1952}), observational techniques prior to the early-1990s were largely incapable of detecting Jupiter-mass companions orbiting nearby stars. The discovery of 51 Pegasi b (\citealt{MayorQueloz1995}) and subsequent close-orbiting hot Jupiters (e.g. \citealt{ButlerMarcy1997}; \citealt{Butler1999}; \citealt{Charbonneau2000}; \citealt{HenryMarcy2000}), brought a paradigm shift to our understanding of planetary systems beyond the Solar System. Within the current framework of planet formation, giant planets are expected to predominantly form beyond the water-ice line, where conditions are conducive to rapid core assembly and subsequent gas accretion (e.g. \citealt{pollack_formation_1996}; \citealt{Lecar2006}; \citealt{KennedyKenyon2008}). This picture is empirically supported by an increase in the occurrence rate of giant planets at wider separations (\citealt{cumming_keck_2008}; \citealt{Fernandes2019}; \citealt{Wittenmyer2020}; \citealt{fulton_California_2021}). Hot Jupiters (with orbital periods $\lesssim$10~days) are intrinsically rare, with occurrence rates of $\sim$0.5--1\% around single Sun-like stars (\citealt{Marcy2005}; \citealt{Mayor2011}; \citealt{Wright2012}; \citealt{Fressin2013}; \citealt{Petigura2018}; \citealt{BeleznayKunimoto2022}). Their existence presents a longstanding challenge in explaining how these Jovian-like planets arrived at their present orbital separations (\citealt{Dawson2018}; \citealt{Fortney2021}).

Several mechanisms have been proposed to explain the origins of hot Jupiters, including inward migration driven by early interactions with the protoplanetary disk (\citealt{GoldreichTremaine1980}; \citealt{LinBodenheimer1996}; \citealt{KleyNelson2012}), high-eccentricity tidal migration induced by processes such as planet--planet scattering (\citealt{rasioford1996}; \citealt{chatterjee2008}) or von Zeipel--Lidov--Kozai (ZLK) oscillations with an outer companion (\citealt{Kozai1962}; \citealt{Lidov1962}; \citealt{Eggleton2001}; \citealt{fabrycky_shrinking_2007}; \citealt{Wu2007}; \citealt{naoz_hot_2011}; \citealt{ItoOhtsuka_Zeipel2019}), and, in some cases, in situ formation (\citealt{Batygin2016}; \citealt{Boley2016}). In the case of high-eccentricity tidal migration, a second companion is necessary to excite the giant planet’s orbit and drive its inward migration.

Early studies revealed that many planet-hosting stars also possess stellar binary companions. This realization motivated a series of high-resolution imaging surveys targeting planet-hosting stars (\citealt{Luhman2002};  \citealt{Patience2002}; \citealt{Eggenberger2007}; \citealt{Ginski2012}; \citealt{Ginski2016}), including those discovered with \textit{Kepler} (\citealt{Adams2012}; \citealt{Dressing2014}; \citealt{JiWang2014}; \citealt{Law2014}; \citealt{Kraus2016}), and \textit{TESS} (\citealt{Ziegler2020}; \citealt{Howell2021}; \citealt{Ziegler2021}; \citealt{Matson2025}; \citealt{Zhang2024}; \citealt{Zhang2026}), as well as stars in the solar neighborhood (\citealt{Hirsch2021}). Similar studies focusing on transiting hot Jupiter systems, albeit with smaller sample sizes, identified several stellar companions (\citealt{Daemgen2009}; \citealt{Bergfors2013}; \citealt{Faedi2013}; \citealt{JiWang2015}) and provided early evidence that the multiplicity rates of these planet-hosting stars are comparable to those of field stars, at least for companions with separations beyond $\sim$50 AU. For instance, \citet{EvansSouthworth2016} report a binary fraction of $38^{+17}_{-13}\%$ among hot Jupiter host stars for companions with separations $a > 200$ AU, while \citet{Ngo2016} find a multiplicity rate of $47^{+7}_{-7}\%$ for companions with separations between 50--2000 AU. These values are consistent with the $\approx$50\% multiplicity fraction observed for for Sun-like stars in the field (\citealt{Raghavan2010}; \citealt{DucheneKraus2013}; \citealt{Offner2023}).

However, these surveys were primarily designed to measure the binary multiplicity fraction in systems hosting giant planets. While useful for determining projected separations and general properties of the stellar companions, the orbital properties and viewing geometry of the planet-hosting binaries remain largely unconstrained. As a result, it is unclear whether stellar companions play a meaningful role in the formation or migration of some hot Jupiters.

The stellar companions targeted in hot Jupiter-hosting systems predominantly reside at separations $>100$ AU and therefore have orbital periods of $\gtrsim 10^3$ yr. Most imaging surveys obtained only two astrometric measurements separated by about one year to establish common proper motion. As a result, only a small fraction of each companion's orbit was sampled, making it difficult to constrain their orbital parameters. Long-baseline relative astrometry combined with absolute astrometry and radial velocities can be used to precisely determine the orbits of close-in ($\lesssim$ 500 AU) binary companions, characterize their orbital architectures, and assess their dynamical influence. For such companions, extending the observational baseline is often more valuable than increasing the cadence of observations, as even modest amounts of additional orbital coverage can substantially improve orbital constraints. New astrometric measurements therefore provide critical leverage by extending the time baseline and refining the companion orbits.

\begin{deluxetable*}{lcccccc}
\tablecaption{New Keck/NIRC2 observations from this work\label{tab:new_nirc2}}
\tablehead{
\colhead{Target} &
\colhead{UT Date} &
\colhead{Filter} &
\colhead{Array} &
\colhead{$T_{\mathrm{int}}$ (s)\tablenotemark{b}} &
\colhead{$N_{\mathrm{frames}}$} &
\colhead{$N_{\mathrm{coadds}}$}
}
\startdata
WASP-2             & 2022-07-12 & $K_s$ & $1024\times1024$ & 0.4        & 9  & 10 \\
                   & 2025-06-30 & $K_p$  & $1024\times1024$ & 0.35       & 10 & 80 \\
                   & 2025-09-30 & $K_s$ & $512\times512$   & 0.5 & 20 & 10 \\
\hline
WASP-76            & 2022-09-17 & $K_s$  & $512\times512$   & 0.15       & 22 & 10 \\
                   & 2025-06-30 & $K_p$ & $512\times512$   & 0.1        & 10 & 80 \\
                   & 2025-09-30 & Fe~II         & $512\times512$   & 0.5        & 30 & 10 \\
\hline
WASP-11 / HAT-P-10 & 2025-09-30 & $K_s$  & $512\times512$   & 0.5        & 30 & 10 \\
\hline
HAT-P-14           & 2022-07-12 & $K_s$  & $512\times512$   & 0.25 & 9 & 10 \\
                   & 2025-06-30 & $K_p$ & $512\times512$   & 0.15       & 20 & 80 \\
\hline
HAT-P-16           & 2022-07-12 & $K_s$ & $512\times512$   & 0.5        & 8  & 10 \\
                   & 2025-09-30 & $K_s$ & $512\times512$   & 0.32       & 30 & 10 \\
\hline
$\tau$ Boo         & 2002-05-01 & $K$+cor600\tablenotemark{a} & $1024\times1024$ & 1.45 & 3  & 5 \\
                   & 2022-07-12 & $K_s$+cor600                 & $512\times512$   & 0.05 & 10 & 10 \\
\enddata
\tablenotetext{a}{``cor600'' refers to the 600 mas-diameter focal-plane coronagraph.}
\tablenotetext{b}{$T_{\mathrm{int}}$ (s) is the integration per coadd.}
\end{deluxetable*}

Over a decade has passed since a large number of efforts were undertaken to search for stellar companions in systems hosting gas giants with AO and speckle imaging (\citealt{Knutson2014}; \citealt{Ngo2015}; \citealt{Piskorz2015}; \citealt{JiWang2015}; \citealt{EvansSouthworth2016};  \citealt{Ngo2016}). Given the temporal baseline established by these studies, there is a unique opportunity to leverage new high-resolution imaging to better constrain the orbits of these stellar companions. In this work, we aim to investigate whether the orbital properties of outer binary companions in systems with inner hot Jupiters play a decisive role in hot Jupiter formation and migration. By combining previously published and new relative astrometry, we establish a time baseline that enables substantially improved orbital constraints for stellar binaries in hot Jupiter--hosting systems. 

This paper is organized as follows. In Section \ref{sec:Target_Sample} we discuss our survey design and target sample. In Section \ref{sec:Data_Reduction} we discuss the observations used for the systems considered in this study.  In Sections \ref{sec:Orbit_fits} and \ref{sec:Results} we present our orbit fitting procedures and contextualize results. Implications for giant planet migration in systems with binaries and comparisons to different binary populations are discussed in Section \ref{sec:Discussion}. Finally, we summarize conclusions and forthcoming work in Section \ref{sec:Conclusion}.

\begin{figure*}
    \centering
    \includegraphics[width=\linewidth]{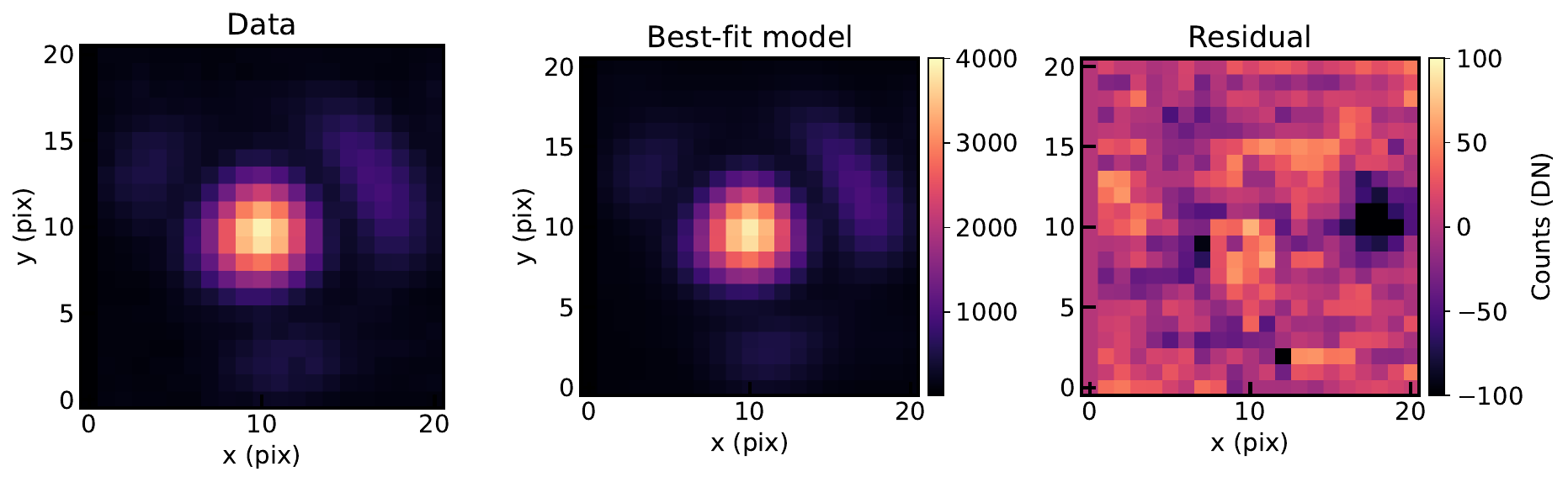}
    \caption{An example relative astrometry fit of a single WASP-2 NIRC2 frame from UT 2025-09-30 ($K_s$ filter). The stellar companion is shown on the left, the model (based on the empirical PSF of the primary star) on the middle, and residuals on the right.}
    \label{fig:nirc2_fit}
\end{figure*}

\section{Target Sample}\label{sec:Target_Sample}

In this study, we introduce the Tracing the Influence of Distant companions on Exoplanetary Systems (TIDES) survey. We present our observational setup, analysis framework, and methodology using six systems with previous Keck/NIRC2 imaging and new follow-up observations obtained over a decade later, enabling improved constraints on their orbital eccentricities and inclinations. The full sample of hot Jupiter-binary hosting stars are described in further detail below.

This sample originates from early discoveries and surveys targeting binary companions in systems with inner hot Jupiters orbiting the primary star (e.g. \citealt{Eggenberger2007}; \citealt{Knutson2014}; \citealt{Ngo2015}; \citealt{Wollert2015}; \citealt{Ginski2016}; \citealt{Ngo2016}). We first select systems hosting confirmed hot Jupiters with orbital periods of $\leq$ 10 days ($\lesssim$ 0.1 AU) and either $m_p \sin i = 0.3$--13~$M_{\mathrm{Jup}}$ or radii $> 8\,R_{\oplus}$, as identified through the NASA Exoplanet Archive (\citealt{Akeson2013}). We first isolate systems with confirmed co-moving stellar companions in S-type binary configurations, in which the planet orbits one of the stellar components. We then assemble previously published relative astrometry or long-baseline radial velocity measurements of the host star. Systems are retained only if the companion was visually resolved and had existing high-resolution imaging measurements (e.g., AO or speckle). We remove any systems with projected separations beyond 500 AU. Although such widely separated binaries may play a role in giant planet migration, their large separations result in poorly constrained orbital arcs, limiting their usefulness for this analysis. Finally, we exclude systems with more than one stellar companion to avoid complications in orbit fitting and astrometric analysis, and because higher-order multiple systems (e.g., triples or quadruples) can have more complicated dynamical impacts on planetary system formation and evolution (e.g. \citealt{Yang2025}; \citealt{Liu_Lu_2026}). Altogether, our remaining sample comprises several dozen transiting and non-transiting hot Jupiters in binary star systems for which measurements of the binary orbital eccentricity and inclination are feasible. In this initial analysis we focus on six of the highest-priority systems that have both previous Keck/NIRC2 astrometry and new NIRC2 imaging from our follow-up observations which together span a sufficiently long astrometric baseline to enable constraints on the binary orbital elements.

\section{Observations and Data Reduction}\label{sec:Data_Reduction} 

\subsection{Keck/NIRC2 imaging}\label{sec:Reduced_NIRC2_Imaging} 

In this work, we collect new NIRC2 imaging of six systems, $\tau$~Boo, WASP-2, WASP-11 (=HAT-P-10), WASP-76, HAT-P-14, and HAT-P-16, between 2022 and 2025 with natural guide star adaptive optics at Keck-II. For consistency, we also re-reduce all archival NIRC2 data for these systems with the same pipeline. All observations analyzed here are obtained in position angle mode---in which the AO rotator maintains a fixed sky orientation on the NIRC2 detector---and most are obtained in the $K_s$ or $K_p$ filters (see Table~\ref{tab:new_nirc2}). We apply several pre-processing steps including flat-field correction, bad-pixel removal, and correction of geometric distortions. For data taken before UT 2015 April 13, we apply the solution from \citet{Yelda2010} and for data taken after, we apply the solution from \citet{Service2016}. 

To measure the relative astrometry between the stellar companions and primary stars, we analyze each frame independently and define a likelihood by modeling the companion's point spread function (PSF) as a scaled copy of the empirical primary star PSF in the same frame. The residuals between data and model are calculated over a 21-by-21 pixel square cutout, and the total log likelihood is the sum over all frames. The parameters fitted are the companion offset in detector coordinates, $(\Delta x,\Delta y)$. We solve for the companion flux and local background flux in each frame analytically at each likelihood evaluation. The posteriors are sampled using \texttt{emcee} \citep{Foreman-Mackey2013} with 1000 steps (300 burn-in) and 32 walkers. We manually inspect the chains for convergence. The posterior samples are converted to projected separation ($\rho$) and position angle ($\theta$) values using the corresponding NIRC2 plate scale and north angle values for the given observing date (i.e. \citealt{Yelda2010} or \citealt{Service2016}). The final reported uncertainties include the statistical uncertainty from the fit, the uncertainty in the NIRC2 plate scale and north angle, and a 1~mas distortion uncertainty recommended by \citet{Service2016}. An example fit can be found in Figure~\ref{fig:nirc2_fit}.

\begin{figure*}
    \centering
    \includegraphics[width=\linewidth]{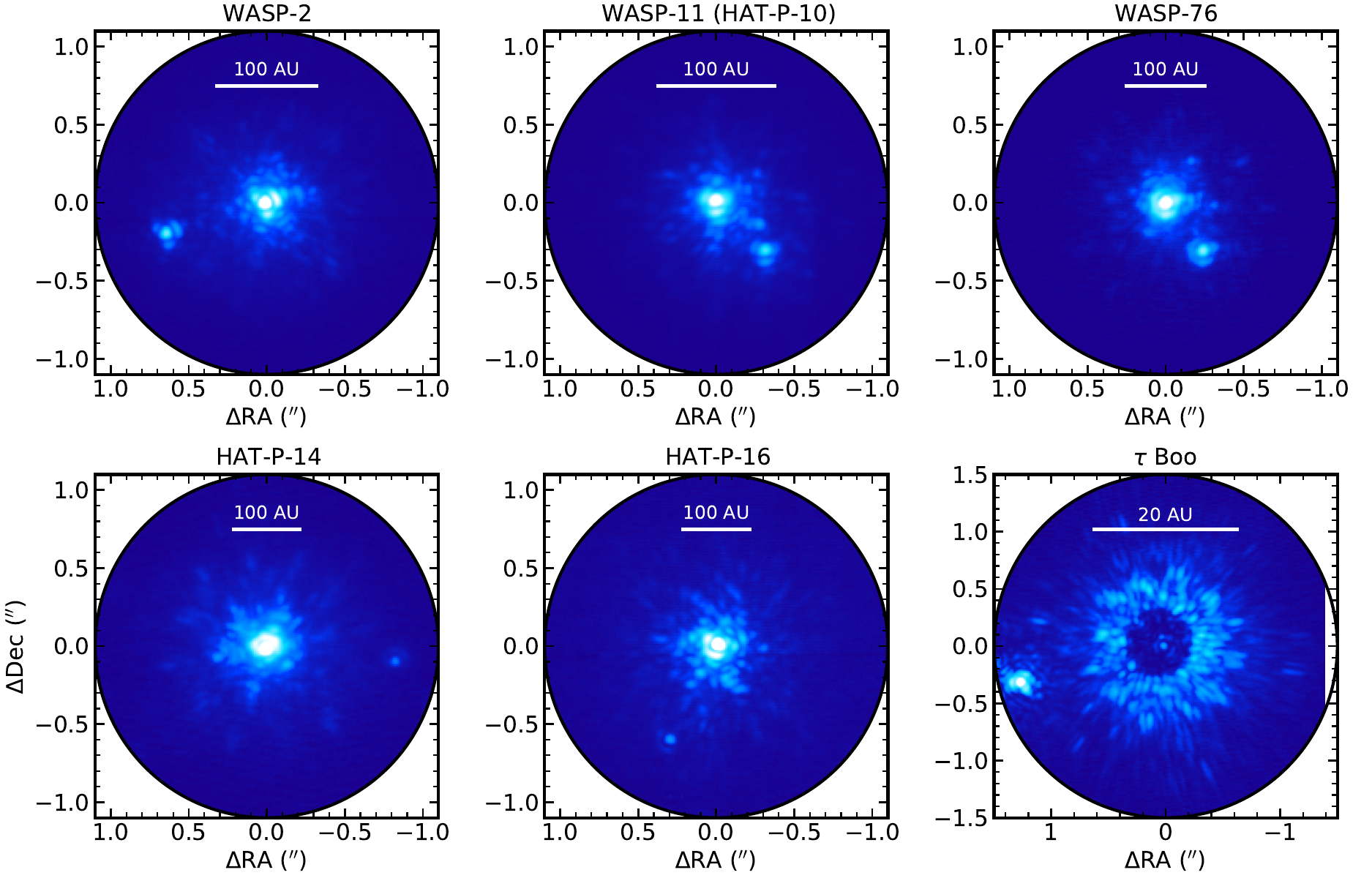}
    \caption{Keck/NIRC2 images of WASP-2, WASP-11, WASP-76, HAT-P-14, HAT-P-16, and $\tau$ Boo all taken in $K_s$ or $K_p$ bands. Each image is oriented such that north is up and and east is to the left. The observations of $\tau$ Boo were obtained in July 2022, those of HAT-P-14 and WASP-76 were obtained in June 2025, and those of WASP-2, WASP-11, and HAT-P-16 were obtained in September 2025.} 
    \label{fig:nirc2_images}
\end{figure*}

For the 2012-2016 archival epochs, we checked that the derived astrometry from this work agrees well ($<0.5\sigma$) with the astrometry published in \citet{Ngo2015} and \citet{Ngo2016}. We provide Keck/NIRC2 images of all six systems in Figure \ref{fig:nirc2_images}. For several systems, we also include published astrometry from VLT/SPHERE, WIYN/NESSI, and other facilities.  A compilation of relative astrometry for all six systems are provided in Table \ref{tab:combined_astrometry}.

The two observations of $\tau$ Boo were carried out behind the 600-mas diameter partly opaque coronagraph.  Astrometry was determined following the procedure described in \citet{Bowler2018}.  Briefly, relative astrometry of $\tau$ Boo A behind the coronagraph and $\tau$ Boo B was carried out for each frame using centroid positions. Uncertainties take into account errors from the north alignment, plate scale, and systematics associated with the residual distortion solution.  In addition to our 2022 observations of $\tau$ Boo, we also present an early epoch of NIRC2 imaging from May 2002 acquired soon after the facility adaptive optics system was made available at Keck II (\citealt{Wizinowich2000}). This system therefore benefits from two decades of adaptive optics imaging baseline with the same instrument.

\subsection{Keck/KPIC spectroscopy}\label{sec:Keck_KPIC_Spectra} 
We observed WASP-2 with the Keck Planet Imager and Characterizer (KPIC) \citep{delorme_Keck_2021a, Jovanovic2025} on UT 2025-06-30. KPIC uses a fiber injection unit to link the Keck-II adaptive optics system and the downstream spectrograph, NIRSPEC (\citealt{Martin2018}; \citealt{lopez_Characterization_2020}). We collected $K$-band spectra at $R\sim35,000$ for both the primary star and the stellar companion with KPIC, which provides a relative RV measurement. During the observation, we also observed a telluric standard star (HIP 61960) and an RV standard star (HIP 62944) for calibration purposes (see \citealt{wang_Detection_2021}). Previous studies have shown that relative RVs can significantly improve orbital constraints when the imaging data only cover a fraction of the orbital period \citep{Xuan2024, YZhang2024, Zhang2026}.

We reduce the data using the public KPIC pipeline\footnote{\url{https://github.com/kpicteam/kpic_pipeline}} to extract 1D, wavelength-calibrated, spectra. For details, see \citet{Xuan2022}. In brief, the main data reduction steps are (1) instrumental background subtraction, (2) trace-finding and measuring the trace widths using the telluric star data, (3) optimal extraction \citep{Horne1986}, and (4) deriving the wavelength solution using the M giant spectra taken on the same night. For the analysis, we use three spectral orders (NIRSPEC orders 31-33) from 2.29-2.49~$\mu$m, which have the most reliable wavelength solution and contain the strongest spectral lines from the stars. 

We fit the KPIC spectra of WASP-2A and WASP-2B using BT-Settl models \citep{Allard2012} to measure the relative RV between the two stars. The grid of BT-Settl models we use covers $3000-6000~$K in \Teff and $3.5-5.5$ in \logg, with 100~$K$ spacing in \Teff and 0.5 dex spacing in \logg. We follow the forward model framework described in \citet{Xuan2024} to fit the data. The fitted parameters are RV, \vsini, \Teff, \logg, flux scales (one per order), and an error inflation term. The continuum is subtracted using a 100-pixel box median before computing the residuals and log likelihood. The posterior is sampled with the nested sampling package \texttt{DYNESTY} \citep{speagle_DYNESTY_2020} using 500 live points. The model fits to the KPIC spectra of WASP-2A and WASP-2B are shown in Appendix \ref{sec:kpic_spectra}, and from the RV measured for each component, we find a relative RV (defined as companion RV - primary RV) of $+0.75\pm0.15~$km/s on UT 2025-06-30.

\begin{figure*}
\begin{center}
{\includegraphics[width=\linewidth]{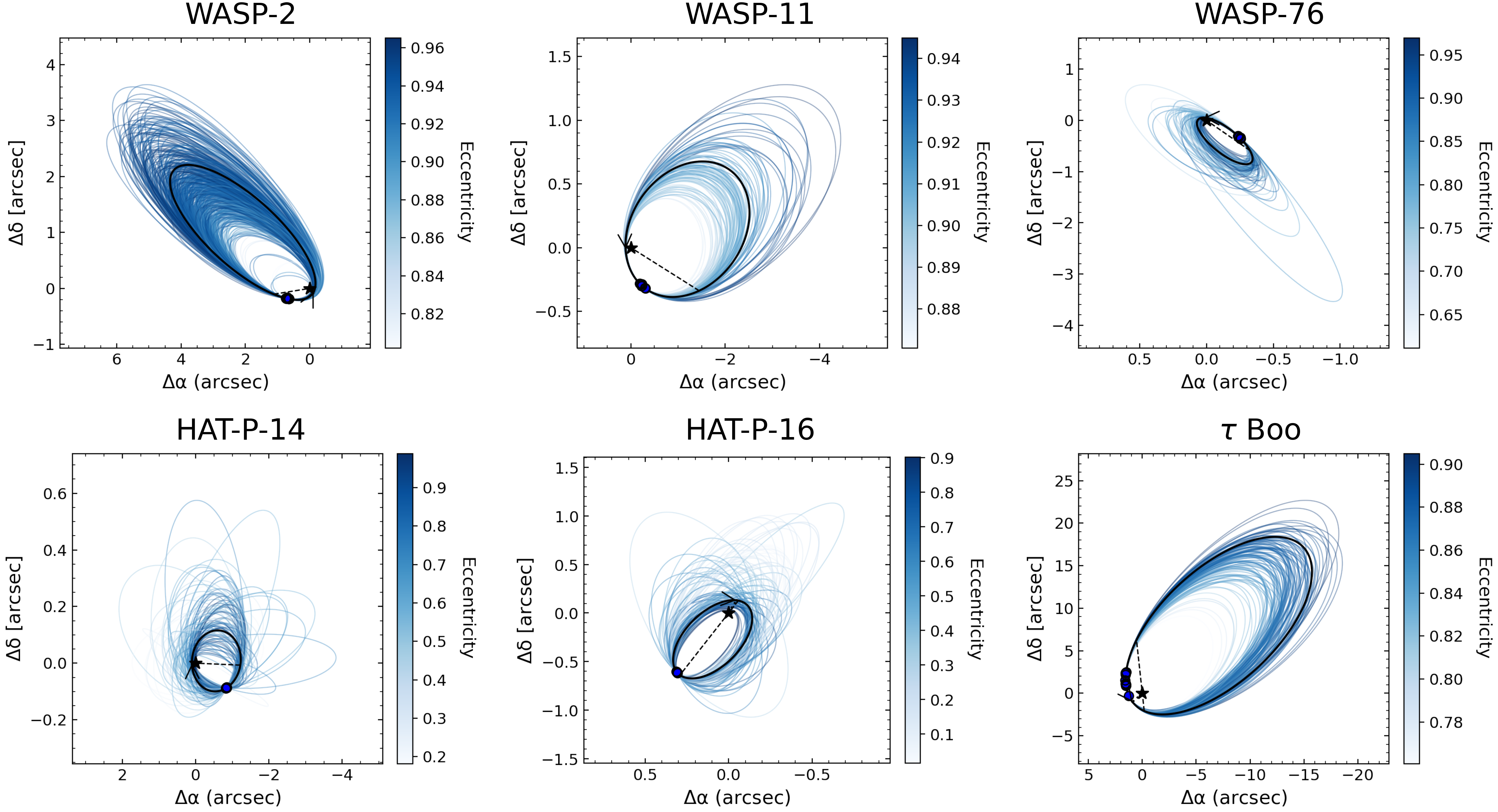}}
\caption{Astrometric orbit fits for WASP-2, WASP-11, WASP-76, HAT-P-14, HAT-P-16, and $\tau$ Boo. The solid black line denotes the best-fit orbit from the MCMC chain. 50 random draws from the posterior orbit fits are colored by the eccentricity of the binary. The dotted lines connect the host star to the periastron passages, while the arrows indicate the direction of motion of the binary companion.} 
\label{fig:Astrometric_Orbit_Fits}
\end{center}
\end{figure*}

\subsection{Radial Velocity Trends}\label{sec:HIRES_RV_Trends} 
Radial accelerations—manifested as long-term drifts in RV measurements over extended baselines—provide constraints on the partial orbits of long-period companions (\citealt{Torres1999}). These RV trends provide complementary constraints to relative and absolute astrometry by probing the line-of-sight motion of the host star induced by a companion. In this analysis, we include RV data only when the measurements exhibit a long-term trend consistent with the binary companion. We compiled Lick/Hamilton (\citealt{Fischer2014}), HARPS-N (\citealt{Borsa2015}) and Keck/HIRES (\citealt{rosenthal_California_2021}) RVs for $\tau$ Boo, SOPHIE (\citealt{CollierCameron2007}) and HARPS (\citealt{Husnoo2012}) RVs for WASP-2A, CORALIE and SOPHIE RVs reported in \citet{West2016} for WASP-76A, and HIRES RVs for HAT-P-14, HAT-P-16, and WASP-11 (\citealt{Knutson2014}). Following the setup described in \citet{Morgan2025}, we use \texttt{RadVel} (\citealt{Fulton2018}) and \texttt{emcee} (\citealt{Foreman-Mackey2013}) to model the RVs and re-fit the Keplerian orbits of the HJs. 

We utilize the standard basis for efficient posterior sampling using orbital period $P$, time of inferior conjunction $T_{C}$, semi-amplitude $K$, and parameterized forms of eccentricity ($e$) and argument of periastron ($\omega$), $\sqrt{e}$ $\cos \omega$ and $\sqrt{e}$ $\sin \omega$.\footnote{To focus on the acceleration of the host star rather than the orbit of the inner hot Jupiter, we assume the close-in giant planet has been completely circularized and fix the eccentricity and argument of periastron by setting $\sqrt{e} \cos \omega$ and $\sqrt{e} \sin \omega$ to 0 for $\tau$ Boo b and WASP-11 b.} We also fit for jitter ($\sigma_i$, with one per instrument $i$), instrumental offsets ($\gamma_i$, with one per instrument $i$), a linear acceleration ($\dot{\gamma}$), and quadratic acceleration term ($\ddot{\gamma}$). To test if an acceleration is statistically preferred in each system, we re-fit each dataset three times: once with with a single-planet Keplerian orbit and no RV trend in the model, a single planet plus a linear trend (a radial acceleration), and a single planet plus curvature (changes in the radial acceleration). We then use the Bayesian Information Criterion ($\Delta$BIC; \citealt{Trotta2007}) to identify the statistically favored model, requiring $\Delta$BIC $> 10$ relative to the next-best, more complex model.

The Hamilton Spectrograph underwent several CCD upgrades, with each detector configuration named for the dewar in which the CCD was mounted: Hamilton2, Hamilton13, Hamilton6, and Hamilton8 (see \citealt{Fischer2014}). Based on 247 observations of $\tau$~Ceti obtained between September 1987 and December 1997, the median RV uncertainty decreased from 9.6~m~s$^{-1}$ before the March 1994 detector upgrade to 2.4~m~s$^{-1}$ afterward. We therefore treat the Hamilton2 observations (June 1987--May 1994) as an independent instrument in our global fit. Using RV standard stars with more than 100 observations obtained between 1995 and 2011, \citet{Fischer2014} found that the zero-point offset between the Hamilton13 and Hamilton6 configurations is smaller than the typical RV measurement uncertainties. We therefore combine the Hamilton13 and Hamilton6 measurements (February 1995--March 2002) into a single RV data set following the procedure in \citet{Morgan2026b}. In contrast, Hamilton8 exhibits a zero-point offset of 13.1~m~s$^{-1}$. Although \citet{Fischer2014} corrected for this offset in their published RVs, we conservatively treat Hamilton8 as an independent instrument in the global fit. For $\tau$~Boo, \citet{Borsa2015} measure independent, evolving long-term RV trends in the Hamilton and HARPS-N datasets, with slopes ranging from $\approx$ --10 to --40~m $\text{s}^{-1}$ yr$^{-1}$. For our combined fit of the Hamilton, HARPS-N, and HIRES RVs for $\tau$ Boo we find curvature with values of --32.62$^{+1.18}_{-1.16}$ m $\text{s}^{-1}$ yr$^{-1}$ and --0.83$^{+0.07}_{-0.06}$ m s$^{-1}$ yr$^{-2}$ (see Appendix \ref{sec:rv_fit}; Figure \ref{fig:Tau_Boo_HAT_P_10_RV_Fits}).

We find that a linear trend is favored for WASP-11 and curvature is favored for the $\tau$~Boo dataset, with $\Delta\mathrm{BIC} > 10$, and that the magnitude of the inferred trends are consistent with the known stellar companions. Both orbit fits can be found in Appendix \ref{sec:rv_fit}. HIRES RVs reported in \citet{Knutson2014} are used in the fit for WASP-11. They report a linear trend of $\dot{\gamma}$ = --5.11$^{+1.17}_{-1.17}$ m $\text{s}^{-1}$ yr$^{-1}$. We recover a consistent linear trend of $\dot{\gamma}$ = --5.03$^{+0.26}_{-0.25}$ m $\text{s}^{-1}$ yr$^{-1}$ prompting the inclusion of this RV data in the orbit fit of the WASP-11 system. 

\subsection{Astrometric Accelerations}\label{sec:Astrometric_Accelerations} 
A star without massive nearby companions will exhibit uniform motion in the plane of the sky. Any deviation from this motion detected with absolute astrometry arises from the reflex motion induced by a distant companion and is measured as a tangential acceleration. A 25-year baseline of proper motion measurements from \textit{Hipparcos} (\citealt{vanLeeuwen2007}) and \textit{Gaia} EDR3 (\citealt{GaiaCollaboration2022}) enables the detection of accelerations induced by long-period companions. These accelerations are reported as proper motion anomalies in \citet{Kervella2022} or Hipparcos-Gaia accelerations in the \textit{Hipparcos}-\textit{Gaia} EDR3 Catalog of Accelerations (HGCA; \citealt{brandt_hgca_2021a}). $\tau$~Boo is the only system in this study with proper motion measurements from both \textit{Hipparcos} and \textit{Gaia}, as it is a nearby ($d = 15.61$ pc), bright ($G = 4.36$ mag) star. In this analysis, we focus exclusively on the HGCA measurement (\citealt{brandt_hgca_2021a}), which we include only for the $\tau$~Boo orbit fit.

\subsection{Updated Stellar Host and Companion Masses}\label{sec:Updated_Masses} 

All six host stars in our sample have Gaia DR3 parallaxes with fractional uncertainties $<5\%$, consistent with typical bright targets with Gaia-quality astrometry, which yield host-mass uncertainties at the $\lesssim 10\%$ level (\citealt{Serenelli2021}). Each system has a mass estimate from the TESS Input Catalog (\citealt{Paegert2021}), and five of the six also have Gaia DR3 Final Luminosity Age and Mass Estimator (FLAME) masses. For each host star, we adopt a weighted mean mass and uncertainty using values from \citet{Justesen2019}, \citet{Ngo2015}, \citet{Ngo2016}, the TESS Input Catalog, and Gaia FLAME.\footnote{We note that TIC stellar masses may be less reliable for stars that have evolved off the main sequence, as they are primarily inferred from an empirical $T_{\rm eff}$--mass relation calibrated for main-sequence stars (\citealt{Stassun2019}). The majority of the host stars in this analysis are bright, Sun-like stars, therefore we combine TIC mass estimates with values from discovery papers and the Gaia FLAME catalog.} The updated Gaia DR3 parallaxes and adopted host masses are listed as priors in Tables~\ref{tab:keck-tauboo-wasp2}--\ref{tab:keck-hatp14-hatp16}.

Using the previously published Keck/NIRC2 $K_S$-band photometry and updated Gaia DR3 distances we recompute companion masses and uncertainties using an empirical mass--luminosity--metallicity relation for M dwarfs (\citealt{Mann2019}; \citealt{Mann2025}). These updated companion-mass estimates are adopted as priors in Tables~\ref{tab:keck-tauboo-wasp2}--\ref{tab:keck-hatp14-hatp16}.

\section{Orbit Fitting}\label{sec:Orbit_fits}

\begin{deluxetable*}{cccccccccc}[]
\tabletypesize{\scriptsize}
\tablecaption{Relative Astrometry}\label{tab:combined_astrometry}
\tablehead{
\colhead{System} & \colhead{BJD $-$ 2450000} & \colhead{UT date} & \colhead{Filter} & \colhead{$\rho$ ($''$)} & \colhead{$\sigma_{\rho}$ ($''$)} & \colhead{$\theta$ ($^\circ$)} & \colhead{$\sigma_{\theta}$ ($^\circ$)} & \colhead{Instrument} & \colhead{Reference}
}
\startdata
$\tau$~Boo & 1143.370 & 1998-11-25 & Br-$\gamma$ & 2.87 & 0.03 & 30.8 & 0.3 & Keck-II 10-m/NIRC & \citet{Patience2002} \\
$\tau$~Boo & 1711.400 & 2000-06-15 & $K_s$ & 2.82 & 0.04 & 33.2 & 1.0 & Keck-II 10-m/NIRC & \citet{Patience2002} \\
$\tau$~Boo & 2395.950 & 2002-05-01 & $K$ & 2.650 & 0.011 & 35.4 & 0.1 & Keck-II 10-m/NIRC2 & this work \\
$\tau$~Boo & 4579.230 & 2008-04-22 & SDSS $i$ & 2.181 & 0.011 & 46.26 & 0.42 & Calar Alto 2.2-m/AstraLux & \citet{Ginski2012} \\
$\tau$~Boo & 5653.860 & 2011-04-02 & -- & 1.908 & 0.022 & 53.18 & 0.06 & Carlos Sánchez 1.5-m/FastCam & \citet{Rica2022} \\
$\tau$~Boo & 5688.750 & 2011-05-07 & -- & 1.893 & 0.002 & 54.55 & 0.41 & Carlos Sánchez 1.5-m/FastCam & \citet{Rica2022} \\
$\tau$~Boo & 6010.840 & 2012-03-24 & -- & 1.821 & 0.010 & 55.86 & 0.15 & Carlos Sánchez 1.5-m/FastCam & \citet{Rica2022} \\
$\tau$~Boo & 6021.200 & 2012-04-03 & H$\alpha$ & 1.826 & 0.005 & 56.6 & 0.1 & Starfire 3.5-m & \citet{Drummond2014} \\
$\tau$~Boo & 6067.820 & 2012-05-20 & -- & 1.814 & 0.004 & 56.27 & 0.16 & Carlos Sánchez 1.5-m/FastCam & \citet{Rica2022} \\
$\tau$~Boo & 6123.930 & 2012-07-15 & -- & 1.776 & 0.019 & 57.23 & 0.54 & Carlos Sánchez 1.5-m/FastCam & \citet{Rica2022} \\
$\tau$~Boo & 6396.360 & 2013-04-13 & -- & 1.733 & 0.006 & 59.12 & 0.16 & Carlos Sánchez 1.5-m/FastCam & \citet{Rica2022} \\
$\tau$~Boo & 6418.260 & 2013-05-05 & -- & 1.748 & 0.006 & 60.06 & 0.13 & Carlos Sánchez 1.5-m/FastCam & \citet{Rica2022} \\
$\tau$~Boo & 9772.000 & 2022-07-12 & $K_s$ & 1.283 & 0.003 & 104.7 & 0.2 & Keck-II 10-m/NIRC2 & This work \\
WASP-2 & 4419.5 & 2007-11 & $z^{\prime}$ & 0.757 & 0.001 & 104.7 & 0.3 & Calar Alto 2.2-m/AstraLux & \citet{Daemgen2009} \\
WASP-2 & 4934.500 & 2009-04-13 & $z^{\prime}$ & 0.761 & 0.009 & 103.5 & 0.2 & NTT 3.5-m/AstraLux & \citet{Bergfors2013} \\
WASP-2 & 5133.500 & 2009-10-29 & $z^{\prime}$ & 0.739 & 0.024 & 104.0 & 1.3 & Calar Alto 2.2-m/AstraLux & \citet{Bergfors2013} \\
WASP-2 & 5874.500 & 2011-11-09 & $z^{\prime}$ & 0.744 & 0.013 & 104.6 & 0.7 & Calar Alto 2.2-m/AstraLux & \citet{Bergfors2013} \\
WASP-2 & 6135.520 & 2012-07-27 & $K_p$ & 0.7343 & 0.0014 & 104.40 & 0.11 & Keck-II 10-m/NIRC2 & This work \\
WASP-2 & 6465.530 & 2013-06-22 & $K_s$ & 0.7298 & 0.0014 & 104.54 & 0.11 & Keck-II 10-m/NIRC2 & This work \\
WASP-2 & 6467.500 & 2013-06-25 & $z^{\prime}$ & 0.711 & 0.019 & 105.1 & 0.5 & Calar Alto 2.2-m/AstraLux & \citet{Wollert2015etal} \\
WASP-2 & 6786.770 & 2014-05-10 & $r_{\rm TCI}$ & 0.709 & 0.010 & 103.4 & 0.2 & Danish 1.54-m/TCI & \citet{EvansSouthworth2016} \\
WASP-2 & 7888.500 & 2017-05-15 & $K_s$ & 0.710 & 0.003 & 104.9 & 0.2 & VLT/SPHERE & \citet{Bohn2020} \\
WASP-2 & 9772.490 & 2022-07-12 & $K_s$ & 0.6818 & 0.0015 & 106.09 & 0.12 & Keck-II 10-m/NIRC2 & This work \\
WASP-2 & 10187.710 & 2023-08-31 & -- & 0.6771 & 0.0045 & 106.3 & 0.4 & SOAR 4.1-m/HRCam & \citet{Tokovinin2024} \\
WASP-2 & 10856.490 & 2025-06-30 & $K_p$ & 0.6657 & 0.0015 & 106.53 & 0.12 & Keck-II 10-m/NIRC2 & This work \\
WASP-2 & 10948.280 & 2025-09-30 & $K_s$ & 0.6644 & 0.0015 & 106.54 & 0.12 & Keck-II 10-m/NIRC2 & This work \\
WASP-11 & 5959.740 & 2012-02-02 & $K_p$ & 0.3419 & 0.0014 & 214.09 & 0.24 & Keck-II 10-m/NIRC2 & This work \\
WASP-11 & 6524.140 & 2013-08-19 & $K_p$ & 0.3545 & 0.0014 & 215.58 & 0.23 & Keck-II 10-m/NIRC2 & This work \\
WASP-11 & 6951.500 & 2014-10-21 & $i^\prime/z^\prime$ & 0.374 & 0.013 & 219.75 & 0.79 & Calar Alto 2.2-m/AstraLux & \citet{Wollert2015etal} \\
WASP-11 & 7087.500 & 2015-03-06 & $i^\prime/z^\prime$ & 0.383 & 0.033 & 218.33 & 1.41 & Calar Alto 2.2-m/AstraLux & \citet{Wollert2015etal} \\
WASP-11 & 7645.080 & 2016-09-13 & $K_s$ & 0.3795 & 0.0015 & 217.19 & 0.22 & Keck-II 10-m/NIRC2 & This work \\
WASP-11 & 9980.000 & 2023-02-05 & 832~nm & 0.439 & 0.003 & 223.651 & 0.5 & WIYN 3.5-m/NESSI & M. Everett, ExoFOP \\
WASP-11 & 10948.390 & 2025-09-30 & $K_s$ & 0.4488 & 0.0014 & 224.32 & 0.18 & Keck-II 10-m/NIRC2 & This work \\
WASP-76 & 6889.500 & 2014-08-20 & SDSS $i$ & 0.4438 & 0.0053 & 214.92 & 0.56 & Calar Alto 2.2-m/AstraLux & \citet{Ginski2016} \\
WASP-76 & 6947.500 & 2014-10-21 & $z^\prime$ & 0.425 & 0.012 & 216.90 & 2.93 & Calar Alto 2.2-m/AstraLux & \citet{Wollert2015} \\
WASP-76 & 7214.120 & 2015-07-10 & Br-$\gamma$ & 0.4423 & 0.0014 & 215.49 & 0.19 & Keck-II 10-m/NIRC2 & This work \\
WASP-76 & 7699.500 & 2016-11-07 & $K_s$ & 0.4360 & 0.0030 & 215.90 & 0.40 & VLT/SPHERE & \citet{Bohn2020} \\
WASP-76 & 8358.500 & 2018-08-28 & 832~nm & 0.427 & 0.003 & 216.674 & 0.5 & WIYN 3.5-m/NESSI & M. Everett, ExoFOP \\
WASP-76 & 9839.420 & 2022-09-17 & $K_s$ & 0.4043 & 0.0014 & 216.89 & 0.20 & Keck-II 10-m/NIRC2 & This work \\
WASP-76 & 10856.620 & 2025-06-30 & $K_p$ & 0.3885 & 0.0014 & 217.49 & 0.21 & Keck-II 10-m/NIRC2 & This work \\
WASP-76 & 10948.380 & 2025-09-30 & Fe~II & 0.3874 & 0.0014 & 217.67 & 0.21 & Keck-II 10-m/NIRC2 & This work \\
HAT-P-14 & 6083.520 & 2012-06-05 & $K_p$ & 0.8571 & 0.0014 & 264.11 & 0.10 & Keck-II 10-m/NIRC2 & this work \\
HAT-P-14 & 6377.660 & 2013-03-26 & $K_s$ & 0.8573 & 0.0014 & 264.20 & 0.10 & Keck-II 10-m/NIRC2 & this work \\
HAT-P-14 & 6845.350 & 2014-07-07 & $K_s$ & 0.8554 & 0.0014 & 264.27 & 0.10 & Keck-II 10-m/NIRC2 & this work \\
HAT-P-14 & 9772.330 & 2022-07-12 & $K_s$ & 0.8377 & 0.0015 & 264.03 & 0.10 & Keck-II 10-m/NIRC2 & this work \\
HAT-P-14 & 10856.440 & 2025-06-30 & $K_p$ & 0.8334 & 0.0015 & 263.83 & 0.10 & Keck-II 10-m/NIRC2 & this work \\
HAT-P-16 & 5959.710 & 2012-02-02 & $K_p$ & 0.6910 & 0.0014 & 153.79 & 0.12 & Keck-II 10-m/NIRC2 & this work \\
HAT-P-16 & 6524.070 & 2013-08-19 & $K_s$ & 0.6886 & 0.0014 & 153.64 & 0.12 & Keck-II 10-m/NIRC2 & this work \\
HAT-P-16 & 9773.000 & 2022-07-12 & $K_s$ & 0.6834 & 0.0015 & 153.00 & 0.12 & Keck-II 10-m/NIRC2 & this work \\
HAT-P-16 & 10948.860 & 2025-09-30 & $K_s$ & 0.6807 & 0.0015 & 152.71 & 0.12 & Keck-II 10-m/NIRC2 & this work \\
\enddata
\end{deluxetable*}

We model the orbits of the binaries using the open-source Python package \texttt{orvara} \citep{brandt_orvara_2021} which uses a parallel-tempering MCMC sampler \texttt{emcee} (\citealt{foreman-mackey_emcee:_2013}; \citealt{vousden_Dynamic_2016}). \texttt{orvara} fits Keplerian orbits to combinations of RV, relative astrometry, and absolute astrometry data from the Hipparcos-Gaia Catalog of Accelerations (HGCA; \citealt{brandt_hgca_2021a}). The majority of hot Jupiter-binary host stars are too faint to be included in the HGCA, prompting us to rely primarily on relative astrometry and, where necessary, radial velocity measurements (see Sections \ref{sec:Keck_KPIC_Spectra} and \ref{sec:HIRES_RV_Trends}). Keck/KPIC relative radial velocities are incorporated into the fit for WASP-2, radial velocity trends are included for WASP-11 and $\tau$~Boo, and HGCA absolute astrometry is incorporated for $\tau$~Boo. We use Gaussian priors on the \textit{Gaia} DR3 parallax, the mass of the stellar primary $M_{\mathrm{pri}}$, and the mass of the binary companion $M_{\mathrm{sec}}$. We use log-uniform priors for the semimajor axis $a$ and jitter $\sigma_{\mathrm{jit}}$, $\sin i$ prior for the inclination $i$, and uniform priors for the remaining fitted parameters: $\sqrt{e}\cos\omega$, $\sqrt{e}\sin\omega$, the mean longitude at the reference epoch ($\lambda_{\mathrm{ref}}$, evaluated at 2010.0), and the longitude of the ascending node $\Omega$.

We run each fit using 10 temperatures, 200 walkers, and $10^{5}$ steps, retaining every 50th step. Convergence is assessed from trace and corner plots (\citealt{Foreman-Mackey2016}; \citealt{HoggForeman-Mackey2018}). Joint posterior distributions of model parameters are displayed and can be found in Appendix \ref{sec:Additional_Statistical_Tests}. Each system is fit with all ancillary data (RVs and, when available, absolute astrometry) along with with two versions of relative astrometry: first using all available published astrometry as seen in Table \ref{tab:combined_astrometry}, including the new measurements presented in this work, and second using only our self-consistent set of Keck/NIRC2 astrometry. The motivation for this is to test the impact of observations from several instruments and astrometric measurement pipelines compared to a single instrument with fewer epochs.  On one hand, incorporating measurements from multiple instruments often increases the overall baseline but can introduce instrument-specific systematic biases.  On the other hand, a smaller number of data points and shorter baseline can result in a poorer constraint but minimizes heterogeneous sources. We find that the results are consistent at the $\lesssim 3\sigma$ level when comparing fits using the full dataset with those obtained using only the self-consistently reduced Keck/NIRC2 data. We adopt the orbit fits that incorporate all available data, as shown in Figure~\ref{fig:Astrometric_Orbit_Fits}.\footnote{We also compare our results using the orbit-modeling framework \texttt{Octofitter} \citep{Thompson2023}. The resulting orbit fits are consistent within 1-$\sigma$ with those obtained using \texttt{orvara}, reinforcing our conclusions.} Overall, we find that the astrometric time baseline, orbital phase coverage, and measurement precision have a greater impact on the resulting orbital constraints than the use of a single instrument---in this case NIRC2. 

\begin{figure}
\begin{center}
{\includegraphics[width=\linewidth]{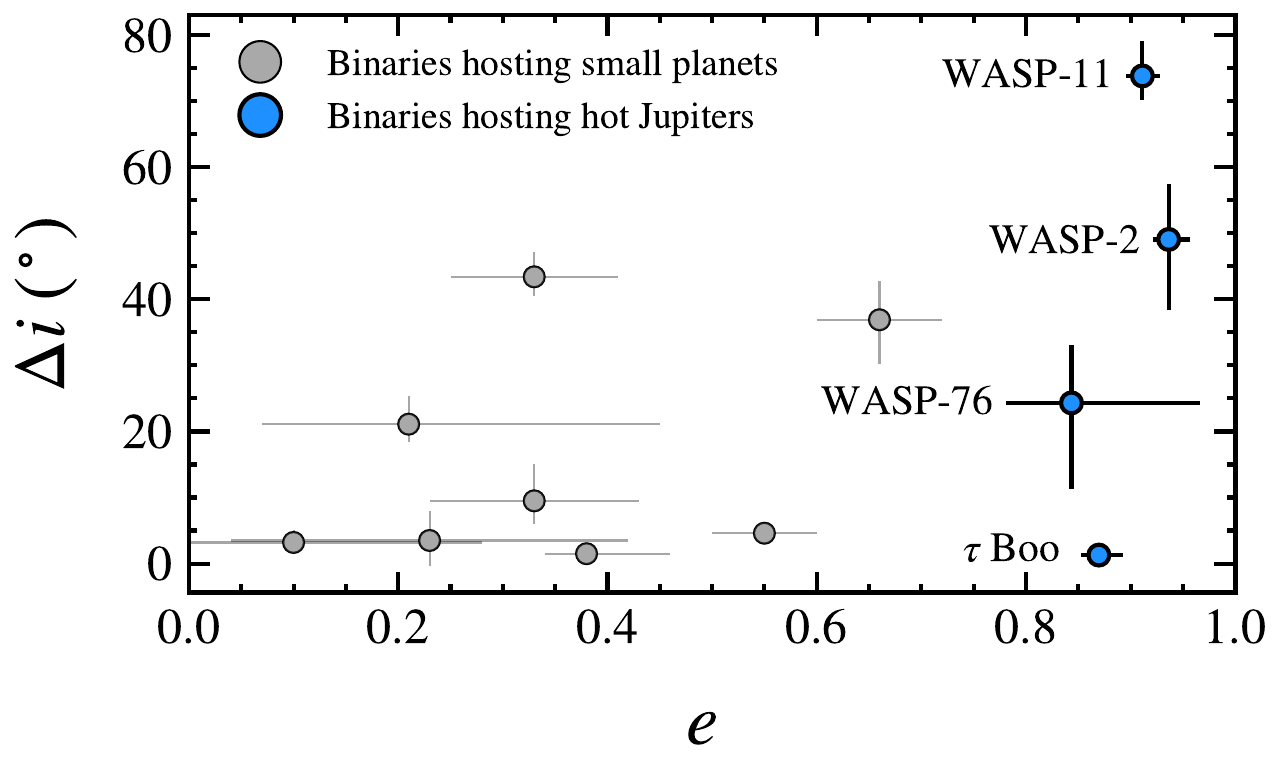}}
\caption{$e$ vs.\ $\Delta i$ for hot Jupiter systems with stellar companion eccentricities constrained at the 15\% level. This includes $\tau$ Boo, WASP-2, WASP-11, and WASP-76. A $\Delta i$ value of $0^\circ$ indicates that the orbital plane of the stellar binary is consistent with being aligned with that of the transiting hot Jupiter, whereas deviations from $0^\circ$ correspond to increasing mutual inclination or orbital misalignment. The grey points represent binaries in systems hosting confirmed inner transiting planets ($R_{\rm p} \lesssim 6\,R_\oplus$) from \citet{Lester2023}, \citet{ZJ_Zhang2023}, and \citet{Zhang2026}. These include TOI-402, TOI-455, TOI-1099, TOI-4399, Kepler-444, KOI 270, KOI 1613, and KOI 4252.} 
\label{fig:e_delta_i}
\end{center}
\end{figure}

\section{Results}\label{sec:Results}

Five of the six planets in this study are transiting, while the sixth, $\tau$~Boo~b, has a measured orbital inclination of $44.5^\circ \pm 1.5^\circ$ derived from high-resolution VLT/CRIRES spectroscopy through detection of the planet's thermal emission (\citealt{Brogi2012}). Transiting planets have orbital inclinations of approximately $90^\circ$, so any deviation of the binary inclination from the planetary inclination---i.e., an edge-on configuration with $i_B \approx i_p \approx 90^\circ$ for transiting systems---provides evidence for misalignment between the planetary and binary orbital planes. Specifically, combining the binary inclination, $i_B$, with the planetary orbital inclination, $i_p$, allows us to place a lower limit on the true planet--binary mutual inclination, $\Delta i = |i_B - i_p|$, because the relative longitude of ascending node, $\Delta \Omega$, is generally unconstrained for transiting planets.

For most binary companions in systems hosting a hot Jupiter, only the projected separation and host–companion mass ratio have been measured, while the orbital eccentricity and mutual inclination remain poorly constrained. The stellar companions in four of the six systems have well-constrained, highly eccentric orbital solutions. We measure eccentricities of $e=0.936^{+0.021}_{-0.015}$ for WASP-2, $e=0.911^{+0.017}_{-0.016}$ for WASP-11, $e=0.844^{+0.123}_{-0.063}$ for WASP-76, and $e=0.870^{+0.023}_{-0.017}$ for $\tau$~Boo. These four systems have an average eccentricity of $e=0.89$. The remaining two of the six systems, HAT-P-14 and HAT-P-16, require additional observations to better constrain their orbital parameters.

 The binary orbital inclinations are constrained for all six systems. We measure $i=40.9^{+10.7}_{-8.4}$$^\circ$ for WASP-2, $i=16.2^{+3.7}_{-5.3}$$^\circ$ for WASP-11, $i=65.7^{+13.0}_{-8.8}$$^\circ$ for WASP-76, $i=43.2^{+1.1}_{-1.1}$$^\circ$ for $\tau$~Boo, $i=102.0^{+6.5}_{-8.7}$$^\circ$ for HAT-P-14, and $i=118.0^{+8.5}_{-11.3}$$^\circ$ for HAT-P-16. The inferred semi-major axes span $118$--$350$~AU, while the companion-to-host mass ratios span $q=0.130$--$0.490$. Together, these orbital solutions (see Table \ref{tab:orbital_params_small_table}) reveal a population of widely separated stellar companions spanning a broad range of orbital inclinations and mass ratios, with the best-constrained orbits strongly favoring high eccentricity values.

 To our knowledge, $\tau$~Boo is the only such hot Jupiter-hosting system with both eccentricity and inclination measurements for the binary companion prior to this work (\citealt{Brogi2012}; \citealt{Justesen2019}). \citet{Justesen2019} incorporated historical relative astrometry for $\tau$~Boo~AB extending back to 1849, providing a $\gtrsim170$~yr baseline and coverage of a substantial fraction of the binary orbit, albeit with comparatively low precision at the earliest epochs. They derived an eccentricity of $0.87^{+0.04}_{-0.03}$ and inclination of $47.2^{+2.7}_{-3.7}$$^\circ$. Our analysis, using a substantially shorter astrometric baseline but higher-precision measurements, yields $e=0.87^{+0.02}_{-0.02}$ and $i=43.2^{+1.1}_{-1.1}$$^\circ$. The eccentricity posteriors are nearly identical, while the inclinations differ by less than $5^\circ$ and are consistent at the 1.2-$\sigma$ level within their uncertainties.

This comparison illustrates the tradeoff between astrometric time baseline, orbital-phase coverage, and measurement precision when constraining long-period binary orbits. A long time baseline can provide substantial orbital leverage even when individual measurements are relatively imprecise, whereas fewer high-precision measurements obtained at particularly informative orbital phases can provide comparable or stronger constraints. In the case of $\tau$~Boo~B, the most recent observations probe the companion as it approaches periastron, where the orbital motion and acceleration become more pronounced. This favorable phase coverage, together with the improved astrometric precision from 10-m telescopes and complementary radial velocity and HGCA constraints, yields an orbital solution consistent with that inferred from more than a century of historical astrometry despite having fewer measurements and a considerably shorter astrometric baseline.

To summarize, we find that four of the six binaries hosting transiting hot Jupiters have highly eccentric orbits ($e \geq 0.8$), and three of these systems exhibit significant planet--binary misalignments ($\Delta i \geq 20^\circ$) as shown in Figure~\ref{fig:e_delta_i}. This supports a scenario in which binaries in systems hosting hot Jupiters tend to be very eccentric and misaligned, indicating that their orbital architectures may provide favorable conditions for dynamical excitation of giant planets.

\begin{deluxetable}{lccccc}
\tablecaption{Orbital Parameters\label{tab:orbital_params_small_table}}
\tabletypesize{\scriptsize}
\tablehead{
\colhead{System} &
\colhead{$e$} &
\colhead{$i$} &
\colhead{$a$} &
\colhead{$q_{\mathrm{bin}}$} &
\colhead{$R_{\mathrm{trunc}}$} \\
\colhead{} &
\colhead{} &
\colhead{($^\circ$)} &
\colhead{(AU)} &
\colhead{(AU)} &
\colhead{(AU)}
}
\startdata
$\tau$ Boo &
$0.870^{+0.023}_{-0.017}$ &
$43.2^{+1.13}_{-1.13}$ &
$197^{+31}_{-30}$ &
$25.6^{+0.8}_{-0.8}$ &
$12^{+5}_{-2}$ \\
WASP-2 &
$0.936^{+0.021}_{-0.015}$ &
$40.9^{+10.7}_{-8.42}$ &
$350^{+68}_{-90}$ &
$23^{+5}_{-7}$ &
$16^{+8}_{-2}$ \\
WASP-11 &
$0.911^{+0.017}_{-0.016}$ &
$16.2^{+3.66}_{-5.25}$ &
$163^{+27}_{-38}$ &
$14^{+1}_{-1}$ &
$8^{+4}_{-1}$ \\
WASP-76 &
$0.844^{+0.123}_{-0.063}$ &
$65.7^{+13.0}_{-8.8}$ &
$118^{+27}_{-39}$ &
$19^{+12}_{-16}$ &
$4.4^{+1.9}_{-0.7}$ \\
HAT-P-14 &
$0.730^{+0.263}_{-0.158}$ &
$102.0^{+6.5}_{-8.7}$ &
$189^{+40}_{-57}$ &
$51^{+43}_{-50}$ &
$12^{+7}_{-3}$ \\
HAT-P-16 &
$0.552^{+0.328}_{-0.193}$ &
$118.0^{+8.5}_{-11.3}$ &
$127^{+22}_{-43}$ &
$58^{+34}_{-50}$ &
$11^{+5}_{-2}$ \\
\enddata
\tablecomments{
We report posterior medians for the eccentricity, inclination, and semimajor axis, with 68.27\% HDI uncertainties. We also report the binary periastron distance, $q_{\mathrm{bin}}=a(1-e)$, and the circumprimary truncation radius--the outer edge of a circumstellar disk, set by the gravitational influence of a binary companion--calculated from Equation~\ref{eq:rtrunc} in Section~\ref{sec:Disk_truncation}. A more complete summary of the orbital parameters and adopted priors is provided in Tables~\ref{tab:keck-tauboo-wasp2}--\ref{tab:keck-hatp14-hatp16}.}
\end{deluxetable}

\section{Discussion}\label{sec:Discussion}
\subsection{Do Close-in Binaries Drive Hot Jupiter Migration?}\label{sec:Binary_HJ_Migration}

Emerging evidence suggests that these binaries are not drawn from the same eccentricity distribution as field binaries at comparable separations. With a sample of thousands of Sun-like binaries from Gaia DR3, \citet{Wu2025} find that systems with $a < 2.5$ AU exhibit a thermal eccentricity distribution that can be well described by a Rayleigh distribution peaking at $e = 0.3$. Using Gaia EDR3 astrometry, \citet{Hwang2022} showed that binaries with separations $\lesssim 100$ AU are approximately uniform in eccentricity, systems at $\sim 10^2$--$10^3$ AU are consistent with a thermal distribution, and binaries at $\gtrsim 10^3$ AU are superthermal, with an excess of highly eccentric orbits, consistent with results from \citet{Tokovinin2020}.\footnote{These are updates from earlier studies which found that close-in field binaries with $a \lesssim 2.5$ AU, as well as longer-period binaries  ($\approx 2.5 < a < 250$ AU), exhibit eccentricities consistent with a roughly uniform distribution (\citealt{Raghavan2010}; \citealt{DucheneKraus2013}).} In contrast, the binaries in our study appear even more strongly superthermal than field binaries at comparable separations, implying a distinct dynamical history.

All four binaries with semi-major axis of $\approx$100--300 AU and $e$ constrained at the $\sigma_e \lesssim 0.1$ level are highly eccentric, with median eccentricities ranging from 0.84 to 0.94. Here, we examine whether the prevalence of such highly eccentric binaries in hot Jupiter systems is statistically unusual, or simply the result of the modest sample of hot Jupiter-hosting systems analyzed. The probability of drawing an eccentricity in the range $0.8 \le e < 1$ is 0.2 for a uniform distribution and 0.36 for a thermal distribution, which follows $\mathcal{P}(e) \propto 2e$ (\citealt{Ambartsumian1937}). The probability that four binaries randomly drawn from the sample of field binaries at comparable separations would each have $e > 0.8$ is therefore $0.2^{4} \approx 0.002$ and $0.36^{4} \approx 0.02$, corresponding to significances of $\approx$ $3.2\sigma$ and $2.4\sigma$, respectively. It is therefore unlikely that all four binaries would exhibit such high eccentricities if they were drawn at random from the eccentricity distribution of field binaries.

\subsection{Disk Truncation and the Formation of Giant Planets}\label{sec:Disk_truncation}
The same binary architectures that can trigger high-eccentricity migration may also alter the conditions under which giant planets form. Eccentric binary orbits can pass close enough at periastron to truncate protoplanetary disks (\citealt{Papaloizou1977}; \citealt{Artymowicz_Lubow_1994}; \citealt{LubowMartin2015}; \citealt{MirandaLai2015}). Using the median posterior orbits in our sample for $\tau$ Boo B, WASP-2 B, WASP-11 B, and WASP-76 B the companion periastron distances are $q_{\rm bin}\approx 15$--$26$ AU. We compute truncation radii using equation 5 from \citet{Tofflemire2024} (adapted from \citealt{Artymowicz_Lubow_1994}),

\begin{equation}
R_{\rm trunc} =
a \,
\frac{0.49\,q^{-2/3}}
{0.6\,q^{-2/3} + \ln\!\left(1 + q^{-1/3}\right)}
\left(b\,e^c + 0.88\,\mu^{0.01}\right),
\label{eq:rtrunc}
\end{equation}

\noindent where $a$ is the binary semimajor axis, $q = M_B/M_A$ is the stellar mass ratio, $e$ is the binary orbital eccentricity, and $\mu = M_B/(M_A + M_B)$ is the secondary-to-total mass ratio. The coefficients $b$ and $c$ depend on $\mu$ and are adopted from the values tabulated in Appendix~C.1 of \citet{Manara2019}. We estimate the truncation radii by drawing Monte Carlo realizations directly from the posterior orbital samples while sampling the tabulated circumprimary coefficients. We find that the median circumstellar truncation radii span $R_{\rm trunc}\approx 4$--$15$ AU. These disks would remain large enough to overlap the canonical giant-planet-forming region, but they would be substantially more compact than disks around single stars.

Truncation may enhance radial drift and viscous evolution in compact disks which can concentrate material at smaller orbital radii (\citealt{Zagaria2021a}; \citealt{Rota2022}), potentially increasing the efficiency of disk-driven migration (\citealt{GoldreichTremaine1980}; \citealt{LinBodenheimer1996}) or enabling in-situ assembly within the dynamically stable circumprimary region (\citealt{Batygin2016}; \citealt{Boley2016}). In this scenario, the binary companion shapes the protoplanetary disk before or during planet formation, and the close-in giant planet may have formed without undergoing an extreme-eccentricity phase (e.g \citealt{KleyNelson2008}). We emphasize that the truncation estimates above use the median posterior values of the binary semimajor axis and eccentricity. The orbit fit posteriors have broad uncertainties, which translate into a range of periastron distances and corresponding disk truncation radii constrained to the 30--50$\%$ level. In addition, both disk truncation and stellar perturbations must be included to accurately model planet formation in S-type binaries. As a result, the inferred disk architecture should be interpreted as illustrative rather than uniquely determined.

These channels can be distinguished observationally. Binary-driven high-eccentricity migration predicts isolated hot Jupiters and large planet--binary mutual inclinations. Recent numerical experiments support this interpretation. For instance, \citet{AvilaBravo2026} show that giant planets in systems with moderate-separation stellar binaries ($\approx$ 50 $<$ $a$ $<$ 500 AU) can be driven to $e \gtrsim 0.8$ when planet--planet scattering and binary-driven secular forcing act together. They find that eccentricities can be excited by inclined and eccentric binary companions and the most extreme outcomes are associated with systems that have been dynamically depleted to a single surviving planet. On the other hand, disk migration or in-situ formation in a truncated disk can accommodate close-in planetary companions and should preserve planet-binary obliquities that more closely trace the primordial disk--binary geometry.

\subsection{Can Secular ZKL Migration Operate in these Systems?}\label{sec:ZKL}
Very eccentric planets, or planets orbiting close to their host stars, experience efficient apsidal precession due to general relativistic effects (\citealt{Einstein1915S}). This rapid precession can inhibit ZKL oscillations, which rely on long-term, coherent secular torques from a distant companion to excite the planet's eccentricity. When the apsidal precession timescale becomes shorter than the ZKL timescale, the orientation of the planetary orbit changes too quickly for the perturbing companion to exert a coherent torque. As a result, the eccentricity excitation is suppressed, preventing the planet from reaching the extreme eccentricities required for high-eccentricity migration and subsequent tidal circularization into a hot Jupiter (\citealt{ford_secular_2000}; \citealt{AdamsLaughlin2006}; \citealt{naoz_eccentric_2016}). We explore whether the orbital architectures presented in this work are conducive to long-term secular interactions.

The initial orbital architectures of the observed hot Jupiters are unknown. We therefore adopt an assumption in which these planets formed at larger separations as cold Jupiters and ask whether ZKL forcing is dynamically capable of overcoming competing apsidal precession.

Following Equation~(1) of \citet{fabrycky_shrinking_2007}, the characteristic ZKL timescale is
\begin{equation}
\tau_{\rm ZKL}=
\frac{2P_{\rm out}^2}{3\pi P_{\rm in}}
\frac{m_1+m_2+m_3}{m_3}
\left(1-e_{\rm out}^2\right)^{3/2},
\end{equation}
where $(m_1,m_2)$ are the host-star/planet masses and $m_3$ is the outer perturber mass. For fixed inner orbit, this implies faster ZKL forcing from a binary perturber for larger $m_3$, smaller $P_{\rm out}$, and larger $e_{\rm out}$.

The competing general relativity apsidal precession (Equation~22 in \citealt{fabrycky_shrinking_2007}) can be expressed as

\begin{equation}
\dot{\omega}_{\rm GR}=
\frac{3G^{3/2}(m_1+m_2)^{3/2}}
{a_{\rm in}^{5/2}c^2(1-e_{\rm in}^2)}.
\end{equation}

\begin{figure}
\begin{center}
{\includegraphics[width=\linewidth]{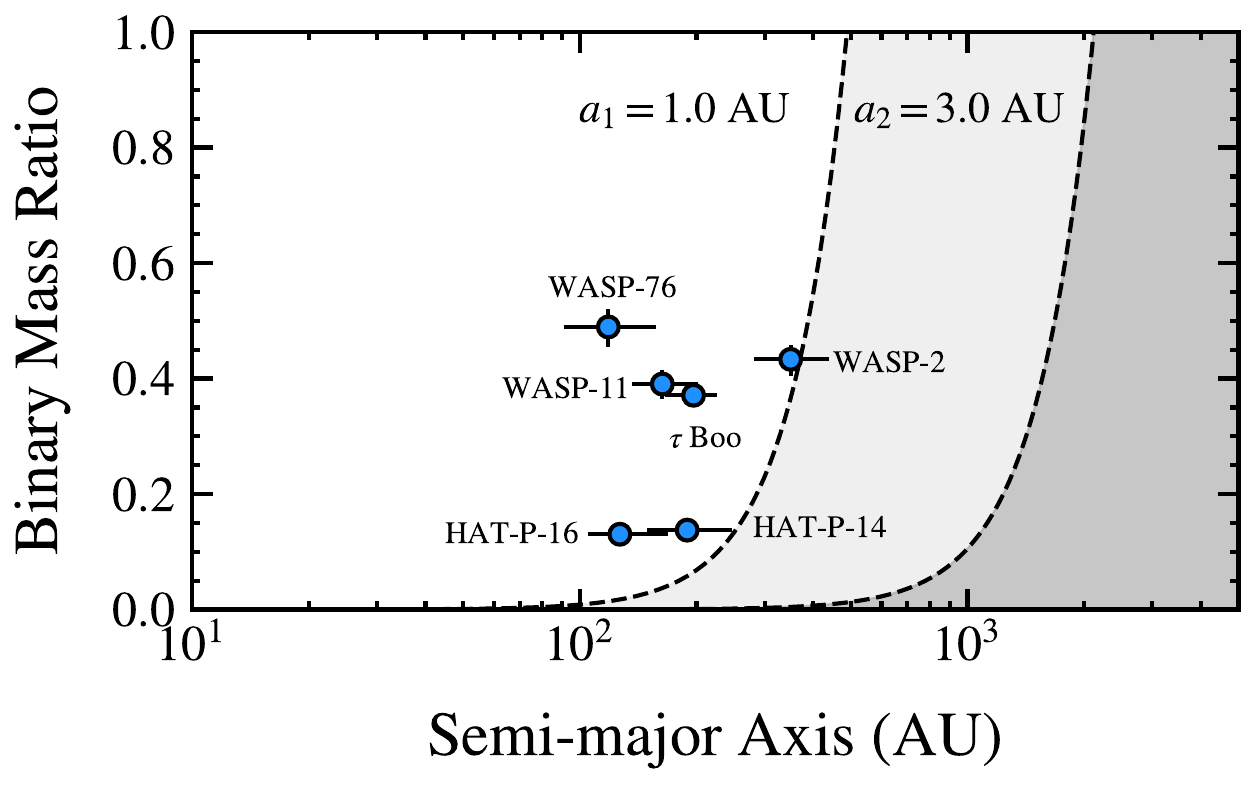}}
\caption{Binary mass ratio in comparison to the semi-major axis for hot Jupiter systems derived in this work. Masses and semi-major axes are derived in this work. The dashed lines represent the the minimum mass needed to induce ZKL oscillations at timescale short enough to overcome general relativistic pericenter precession. We assume initial starting positions of $a_{1}$ = 1 AU or $a_{2}$ = 3 AU for a 1 $M_{\mathrm{Jup}}$ planet, 1 $M_\odot$ host star, and an eccentric companion $e$ = 0.8. Taken at face value, stellar binaries interior to these dashed lines reside in a regime that is favorable to induce ZKL oscillations. } 
\label{fig:mass_ratio_separation}
\end{center}
\end{figure}

\noindent In this context, general relativity precession is a drift of the argument of pericenter of the planet ($\omega$) with time.  Because ZKL eccentricity growth requires a secular torque from the outer companion, rapid additional precession (especially from general relativistic effects, and in some systems tidal or rotational precession) averages out that torque, so efficient excitation requires \(\tau_{\rm ZKL}\dot{\omega}_{\rm GR}\lesssim \mathcal{O}(1)\), whereas \(\tau_{\rm ZKL}\dot{\omega}_{\rm GR}\gg 1\) implies suppression. ZKL forcing must act at least as fast as GR precession, otherwise eccentricity growth is weakened or quenched (\citealt{fabrycky_shrinking_2007}, \citealt{petrovich_steady-state_2015}). As seen in Figure \ref{fig:mass_ratio_separation}, we find that the binaries analyzed in this initial study have masses and separations conducive to ZKL forcing assuming a solar-like mass host star, eccentric outer binary ($e$ = 0.8), and planet forming at 1--3 AU. Four of the binaries have well-constrained eccentricities, whereas HAT-P-14 and HAT-P-16 require additional observations to better constrain their orbital eccentricities. For the purposes of this analysis, we assume $e > 0.8$ for all systems.
 
These assumptions do not take into account the large mutual inclination and long-term dynamical evolution required for these secular interactions. In the quadrupole limit, large-amplitude ZKL cycles require mutual inclinations above $\simeq 39^\circ$, however, eccentric outer companions activate octupole-order dynamics---eccentric ZKL effects---that can broaden the range of effective inclinations and produce chaotic eccentricity growth (\citealt{lithwick_eccentric_2011}; \citealt{naoz_eccentric_2016}). For initially wide giant planets, general-relativistic precession is unlikely to be the dominant barrier to eccentricity excitation; instead, the limiting factor is primarily geometric.

\subsection{Differences in Binary Properties Between Hot Jupiter and Small-Planet Systems}\label{sec:small_transiting_planet_comparison}

Observations combining adaptive optics, high-resolution imaging, and constraints from \textit{Gaia}, \textit{TESS}, and \textit{Kepler} find that small inner transiting planets ($R_{\rm p} \lesssim 6\,R_\oplus$) are preferentially aligned with the orbital planes of low-mass stellar binaries in their systems (\citealt{Dupuy2016}; \citealt{Christian2022}; \citealt{Behmard2022}; \citealt{DupuyKraus2022}; \citealt{Lester2023}; \citealt{ZJ_Zhang2023}; \citealt{Christian2025}). Similarly, \citet{Rice2024} combine \textit{Gaia} DR3 astrometry with stellar obliquity measurements and find that nine exoplanet systems—including both low-mass and giant transiting planets—are consistent with alignment among the host star, the transiting planet, and outer binary companions, as indicated by both spin--orbit and orbit--orbit alignment. On the other hand, \citet{Zhang2026} find a mixture of aligned and misaligned systems among five stellar companions with well-constrained orbits in systems hosting small transiting planets.

One possible explanation for this alignment trend is that a close-in stellar companion can torque and align the protoplanetary disk of the primary star, thereby producing the observed alignment between transiting planets and stellar binaries at the typical ages of field stars (\citealt{Bate2000}; \citealt{LubowOgilvie2000}; \citealt{Batygin2012}; \citealt{Lai2014}; \citealt{Zanazzi2018_discwarp}; \citealt{Offner2023}). Alternatively, the observed planet-binary alignment may instead be primordial in origin. This is supported by observations of some young binary stars hosting protoplanetary disks that are coplanar with the binary orbits; for example, \citet{Tofflemire2024} found evidence that the protoplanetary disk and binary orbit are aligned in the pre-main sequence binary FO Tau AB.

The eccentric and misaligned binary companions in systems hosting transiting hot Jupiters are qualitatively different from those observed in systems with small transiting planets. Binaries in the transiting small planet sample span a range of eccentricities from $\sim$0.1--0.65 but lack eccentricities above $e=0.8$ similar to those observed for the hot Jupiter sample (see Figure \ref{fig:e_delta_i}). On the other hand, the majority of systems in both samples have minimum planet--binary mutual inclinations $\lesssim 43^\circ$ and do not exhibit diverging behavior, with the exception of WASP-11. The difference in eccentricities between these populations suggests that the orbital architectures of binary companions in hot Jupiter hosting systems may play an important role in driving giant planet migration. In the future, a larger sample will allow us to further explore the relationship between binary orientation and inner planetary system architecture.

\subsection{Potential Biases}\label{sec:Biases}
Several observational biases affect the inferred orbital properties of substellar and stellar companions detected with direct imaging. \citet{Bowler2020} and \citet{Bowler2023} highlight an inclination-dependent discovery bias that depends on the inner working angle (IWA)--the minimum angular separation from the host star at which a companion can be detected. This could be the telescope's diffraction limit or an optical element like a coronagraph. They found that companions on circular, face-on orbits near the IWA are more readily detectable than those on more edge-on orbits, as companions on inclined orbits spend a smaller fraction of their orbital period at projected separations beyond the IWA. This bias is most significant for companions whose angular semimajor axes, $\alpha$, are comparable to the IWA, such that $\alpha/\mathrm{IWA} \sim 1$. Systems with $\alpha/\mathrm{IWA} \gtrsim 2$ are less strongly affected by this bias.

Eccentricity introduces an additional selection effect. Companions on eccentric orbits reach larger projected separations near apastron and spend a disproportionate fraction of their orbital period at these separations. As a result, eccentric companions are generally more likely to be detected than companions on near-circular orbits, largely independent of their orbital orientation (e.g. \citealt{DupuyLiu2011}). However, eccentric companions also spend a portion of their orbit inside the IWA, making detectability a function of orbital phase as well as viewing geometry. \citet{Bowler2023} found that companions on more edge-on orbits are recovered at rates that are lower by factors of $\sim$2--3 relative to more face-on companions, with the decline occurring at inclinations of $\sim40^\circ$ for low-eccentricity orbits and $\sim60^\circ$ for high-eccentricity orbits. Similar eccentricity--inclination biases have been discussed by \citet{Oneil2019} and \citet{Ferrer2021} in the context of companions with limited orbital phase coverage.

These biases are especially relevant for the present sample because many hot-Jupiter host stars with imaged stellar companions are relatively distant. The median host-star distance for these six systems is 204 pc, with most systems lying between $\sim$100 and 300 pc; the nearest exception is $\tau$ Boo at 15.61 pc. For more distant systems, a fixed angular IWA corresponds to a larger projected physical separation, limiting sensitivity to companions at small semimajor axes. This introduces an observational bias, whereby facilities with larger IWAs are less sensitive to companions at small projected separations and more susceptible to inclination and eccentricity biases. Most binary companions in hot-Jupiter systems have been discovered with $\sim$2--10-m class facilities, including the 2-m Calar Alto telescope, the 4.1-m SOAR telescope, the 8.2-m VLT, and the 10-m Keck-II telescope. These facilities generally provide sufficient angular resolution to detect low-mass stellar companions at tens of au around typical hot-Jupiter hosts.

The binary companions presented in this work have semimajor axes ranging from 118--350 au. WASP-2 B and WASP-76 B were originally discovered with the 2.2 m Calar Alto telescope in the $i^\prime$ and $z^\prime$ bands, while the companions to WASP-11 B, HAT-P-14 B, and HAT-P-16 B were identified with Keck-II in $K$ band. The discovery of $\tau$ Boo B is attributed to Herschel in 1825 (\citealt{Herschel1826}; \citealt{Mason2001}; \citealt{Justesen2019}) using a 0.5-m telescope. 

 The companions analyzed here have $\alpha/\mathrm{IWA}$ values ranging from $\sim$6--42, assuming an IWA comparable to the diffraction limit, 1.2 $\lambda/D$, placing them well outside the nominal inner working angles of their respective discovery facilities.\footnote{The corresponding values of $\alpha/\mathrm{IWA}$ are 23 for WASP-2 B, 23 for WASP-11 B, 6 for WASP-76 B, 15 for HAT-P-14 B, 10 for HAT-P-16 B, and 42 for $\tau$ Boo B.} Consequently, the companions in our sample occupy a separation regime where detectability is not strongly limited by the IWA of the original observations and both inclination and eccentricity discovery biases are not expected to strongly impact these systems.

\section{Conclusion}\label{sec:Conclusion}
In this work, we introduce the design and initial results of the TIDES survey, which is designed to monitor stellar companion orbits in systems hosting close-in hot Jupiters to assess the impact of stellar binaries on hot Jupiter formation and evolution. Below is a summary of key results from this study.

\begin{itemize}
\setlength{\topsep}{0pt}
\setlength{\partopsep}{0pt}
\setlength{\itemsep}{0pt}
\setlength{\parsep}{0pt}

\item We present new orbital fits for six systems---$\tau$~Boo, WASP-2, WASP-11, WASP-76, HAT-P-14, and HAT-P-16---by combining previously published astrometry with new NIRC2 observations obtained between 2022 and 2025, thereby establishing time baselines spanning more than a decade.

\item Among hot Jupiter hosts in our sample, binaries with eccentricities constrained to the $\sigma_e \lesssim 0.1$ level exhibit high eccentricities ($e \geq 0.8$). Among these, three systems exhibit significant planet--binary misalignments ($\Delta i \geq 20^\circ$) relative to the inner transiting giant planet.

\item These results suggest that the population of binary companions to hot Jupiter host stars is distinct from both field binaries and binaries hosting inner transiting small planets. Our broader sample will enable a more robust assessment of these differences.

\item The orbits of these stellar companions are consistent with influencing hot Jupiter migration through either high-eccentricity migration or participating in early dynamical evolution that ultimately led to their arrival at small separations.

\end{itemize}
Continued orbital monitoring through relative astrometry, time-series absolute astrometry from Gaia DR4, and extended radial velocity baselines will enable precise constraints on the orbits of stellar companions in even more hot Jupiter-hosting systems.

\begin{acknowledgments}

We thank Jingwen Zhang and Lillian Jiang for insightful conversations on the orbital configurations of the binary companions. MM is grateful to Fred Adams for informative discussions on the effects of general relativistic apsidal precession on close-in giant planets.  B.P.B. acknowledges support from the Alfred P. Sloan Foundation.
J.W.X is thankful for support from the Heising-Simons Foundation 51 Pegasi b Fellowship (grant \#2025-5887). A.S acknowledges support from the National Science Foundation Graduate Research Fellowship under Grant No.~2139433.

This work was supported by a NASA Keck PI Data Award, administered by the NASA Exoplanet Science Institute. Data presented herein were obtained at the W. M. Keck Observatory from telescope time allocated to the National Aeronautics and Space Administration through the agency's scientific partnership with the California Institute of Technology and the University of California. Some of the data presented herein were obtained at Keck Observatory, which is a private 501(c)3 non-profit organization operated as a scientific partnership among the California Institute of Technology, the University of California, and the National Aeronautics and Space Administration. The Observatory was made possible by the generous financial support of the W. M. Keck Foundation. 
The authors wish to recognize and acknowledge the very significant cultural role and reverence that the summit of Maunakea has always had within the Native Hawaiian community. We are most fortunate to have the opportunity to conduct observations from this mountain. 

This research has made use of the Washington Double Star Catalog maintained at the U.S. Naval Observatory. This research has made use of the VizieR catalog access tool,
CDS, Strasbourg, France (doi:10.26093/cds/vizier). The
original description of the VizieR service was published in
\citet{Ochsenbein2000}. This work has made use of data from the European Space Agency (ESA) mission {\it Gaia} (\url{https://www.cosmos.esa.int/gaia}), processed by the {\it Gaia} Data Processing and Analysis Consortium (DPAC,
\url{https://www.cosmos.esa.int/web/gaia/dpac/consortium}). Funding for the DPAC has been provided by national institutions, in particular the institutions participating in the {\it Gaia} Multilateral Agreement.
\end{acknowledgments}

\facilities{Keck:II (NIRC2); Keck:II (KPIC)}

\software{\texttt{numpy} \citep{Harris2020}, \texttt{matplotlib}  \citep{Hunter2007}, \texttt{corner} \citep{Foreman-Mackey2016}, \texttt{emcee} \citep{Foreman-Mackey2013}, \texttt{ArviZ} \citep{arviz2019}, \texttt{pandas} \citep{reback2020pandas}; \texttt{Orvara} \citep{brandt_orvara_2021}; \texttt{radvel} \citep{Fulton2018} }


\appendix
\section{Orbit Fits}\label{sec:nirc2_astromtery}

Here, we present the best-fit orbital parameters for the binaries in our sample. The median values, priors, and the upper and lower bounds of the 68.3\% highest density intervals (HDIs) of the posterior probability distributions are reported in Tables~\ref{tab:keck-tauboo-wasp2}--\ref{tab:keck-hatp14-hatp16}.

\begin{longrotatetable}
\begin{deluxetable*}{lcccccc}
\tablecaption{Astrometry Orbit Fits: $\tau$ Boo AB and WASP-2 AB\label{tab:keck-tauboo-wasp2}}
\tabletypesize{\scriptsize}
\tablehead{
\colhead{} & \multicolumn{3}{c}{$\tau$ Boo AB} & \multicolumn{3}{c}{WASP-2 AB} \\
\cline{2-4}\cline{5-7}
\colhead{} & \colhead{Posterior\tablenotemark{a}} & \colhead{95.4\% HDI} & \colhead{Prior} & \colhead{Posterior\tablenotemark{a}} & \colhead{95.4\% HDI} & \colhead{Prior}
}
\startdata
\multicolumn{7}{c}{Fitted Parameters}\\
\hline
$M_{\star}$ ($M_\odot$) & $1.358^{+0.025}_{-0.030}$ & $1.306,\ 1.421$ & $\mathcal{N}(1.350, 0.030)$ & $0.859^{+0.029}_{-0.030}$ & $0.800,\ 0.918$ & $\mathcal{N}(0.840, 0.030)$ \\
$M_{\mathrm{comp}}$ ($M_\odot$) & $0.504^{+0.011}_{-0.011}$ & $0.482,\ 0.527$ & $\mathcal{N}(0.490, 0.020)$ & $0.372^{+0.010}_{-0.010}$ & $0.352,\ 0.392$ & $\mathcal{N}(0.370, 0.010)$ \\
$a$ (AU) & $197^{+31}_{-30}$ & $134,\ 258$ & $1/a$ (log-flat) & $350^{+68}_{-89}$ & $218,\ 543$ & $1/a$ (log-flat) \\
$i$ ($^\circ$) & $43.2^{+1.1}_{-1.1}$ & $40.9,\ 45.5$ & $\sin(i)$, $0<i<180$ & $40.9^{+10.7}_{-8.4}$ & $22.4,\ 56.9$ & $\sin(i)$, $0<i<180$ \\
$\sqrt{e}\cos\omega$ & $-0.554^{+0.053}_{-0.057}$ & $-0.671,\ -0.438$ & $\mathcal{U}(-1, 1)$ & $0.763^{+0.040}_{-0.035}$ & $0.684,\ 0.839$ & $\mathcal{U}(-1, 1)$ \\
$\sqrt{e}\sin\omega$ & $0.748^{+0.044}_{-0.045}$ & $0.661,\ 0.831$ & $\mathcal{U}(-1, 1)$ & $-0.591^{+0.048}_{-0.054}$ & $-0.686,\ -0.485$ & $\mathcal{U}(-1, 1)$ \\
$\Omega$ ($^\circ$) & $4.294^{+2.192}_{-2.068}$ & $0.000,\ 8.380$ & $\mathcal{U}(0,360)$ & $274.828^{+3.580}_{-4.629}$ & $267.911,\ 285.339$ & $\mathcal{U}(0,360)$ \\
$\lambda_{\mathrm{ref}}$\tablenotemark{a} ($^\circ$) & $123.019^{+4.252}_{-3.889}$ & $114.689,\ 131.992$ & $\mathcal{U}(0,360)$ &
$315.147^{+3.434}_{-3.200}$ & $308.165,\ 322.125$ & $\mathcal{U}(0,360)$ \\
Parallax (mas) & $64.080^{+0.100}_{-0.102}$ & $63.881,\ 64.284$ & $\mathcal{N}(64.047, 0.109)$ & $6.587^{+0.026}_{-0.027}$ & $6.535,\ 6.640$ & $\mathcal{N}(6.578, 0.027)$ \\
\hline
\multicolumn{7}{c}{Derived Parameters}\\
\hline
$P$ (yr) & $2020^{+450}_{-509}$ & $1054,\ 2999$ & $\cdots$ & $5900^{+1587}_{-2353}$ & $2732,\ 11341$ & $\cdots$ \\
$e$ & $0.870^{+0.023}_{-0.017}$ & $0.822,\ 0.907$ & $\cdots$ & $0.936^{+0.021}_{-0.015}$ & $0.889,\ 0.967$ & $\cdots$ \\
$\omega$ ($^\circ$) & $126.557^{+4.391}_{-4.183}$ & $117.513,\ 135.217$ &
$\cdots$ & $322.249^{+3.685}_{-3.732}$ & $315.006,\ 329.954$ &
$\cdots$ \\
$T_{\mathrm{p}}$ (BJD$-2450000$) & $-725880^{+186167}_{-164295}$ & $-1083596,\ -373113$ & $\cdots$ & $-2109008^{+857392}_{-581608}$ & $-4090961,\ -945993$ & $\cdots$ \\
$q\,(M_{\mathrm{comp}}/M_{\star})$ & $0.371^{+0.010}_{-0.009}$ & $0.351,\ 0.390$ & $\cdots$ & $0.433^{+0.019}_{-0.019}$ & $0.396,\ 0.472$ & $\cdots$ \\
\enddata
\tablenotetext{a}{The median value is reported along with the upper and lower bounds of the 68.3$\%$ highest density interval regions of the posterior probability distributions about the median.}
\tablenotetext{b}{Mean longitude at the reference epoch of 2010.0.}
\tablecomments{$\mathcal{U}(a,b) = 1/(b-a)$ between $a$ and $b$, and 0 elsewhere. $\mathcal{N}(\mu, \sigma)$ denotes a Gaussian distribution with mean $\mu$ and standard deviation $\sigma$. Stellar host and companion mass priors are discussed in Section \ref{sec:Orbit_fits}.}
\end{deluxetable*}
\end{longrotatetable} 

\begin{longrotatetable}
\begin{deluxetable*}{lcccccc}
\tablecaption{Astrometry Orbit Fits: WASP-11 AB and WASP-76 AB\label{tab:keck-wasp11-wasp76}}
\tabletypesize{\scriptsize}
\tablehead{
\colhead{} & \multicolumn{3}{c}{WASP-11 AB} & \multicolumn{3}{c}{WASP-76 AB} \\
\cline{2-4}\cline{5-7}
\colhead{} & \colhead{Posterior} & \colhead{95.4\% HDI} & \colhead{Prior} & \colhead{Posterior} & \colhead{95.4\% HDI} & \colhead{Prior}
}
\startdata
\multicolumn{7}{c}{Fitted Parameters}\\
\hline
$M_{\star}$ ($M_\odot$) & $0.847^{+0.028}_{-0.029}$ & $0.789,\ 0.904$ & $\mathcal{N}(0.836, 0.030)$ & $1.372^{+0.031}_{-0.030}$ & $1.312,\ 1.431$ & $\mathcal{N}(1.370, 0.030)$ \\
$M_{\mathrm{comp}}$ ($M_\odot$) & $0.331^{+0.010}_{-0.010}$ & $0.311,\ 0.350$ & $\mathcal{N}(0.330, 0.010)$ & $0.672^{+0.031}_{-0.030}$ & $0.611,\ 0.732$ & $\mathcal{N}(0.670, 0.030)$ \\
$a$ (AU) & $163^{+27}_{-38}$ & $108,\ 244$ & $1/a$ (log-flat) & $118^{+27}_{-39}$ & $71,\ 260$ & $1/a$ (log-flat) \\
$i$ ($^\circ$) & $16.2^{+3.7}_{-5.2}$ & $9.5,\ 29.2$ & $\sin(i)$, $0<i<180$ & $65.7^{+13.0}_{-8.8}$ & $26.5,\ 80.7$ & $\sin(i)$, $0<i<180$ \\
$\sqrt{e}\cos\omega$ & $0.899^{+0.068}_{-0.055}$ & $0.655,\ 0.972$ & $\mathcal{U}(-1, 1)$ & $0.780^{+0.199}_{-0.135}$ & $-0.563,\ 0.985$ & $\mathcal{U}(-1, 1)$ \\
$\sqrt{e}\sin\omega$ & $0.278^{+0.349}_{-0.240}$ & $-0.421,\ 0.771$ & $\mathcal{U}(-1, 1)$ & $-0.411^{+0.250}_{-0.379}$ & $-0.908,\ 0.359$ & $\mathcal{U}(-1, 1)$ \\
$\Omega$ ($^\circ$) & $80.053^{+12.833}_{-21.241}$ & $51.496,\ 120.875$ & $\mathcal{U}(0,360)$ & $27.952^{+8.252}_{-9.341}$ & $5.500,\ 211.442$ &
$\mathcal{U}(0,360)$ \\
$\lambda_{\mathrm{ref}}$ ($^\circ$) & $22.327^{+22.717}_{-15.953}$ & $0.000,\ 56.638$ & $\mathcal{U}(0,360)$ & $297.973^{+21.850}_{-19.005}$ & $115.939,\ 355.250$ & $\mathcal{U}(0,360)$ \\
Parallax (mas) & $7.718^{+0.054}_{-0.055}$ & $7.610,\ 7.826$ & $\mathcal{N}(7.700, 0.058)$ & $5.303^{+0.082}_{-0.081}$ & $5.142,\ 5.469$ & $\mathcal{N}(5.290, 0.083)$ \\
\hline
\multicolumn{7}{c}{Derived Parameters}\\
\hline
$P$ (yr) & $1922^{+480}_{-674}$ & $982,\ 3503$ & $\cdots$ & $903^{+316}_{-416}$ & $403,\ 2925$ & $\cdots$ \\
$e$ & $0.911^{+0.017}_{-0.016}$ & $0.878,\ 0.942$ & $\cdots$ & $0.844^{+0.123}_{-0.062}$ & $0.661,\ 0.973$ & $\cdots$ \\
$\omega$ ($^\circ$) & $16.887^{+20.454}_{-18.000}$ &
$0.000,\ 50.597$ & $\cdots$ & $333.013^{+23.807}_{-27.175}$ &
$271.036,\ 360.000$ & $\cdots$ \\
$T_{\mathrm{p}}$ (BJD$-2450000$) & $-5211^{+267}_{-251}$ & $-5750,\ -4719$ & $\cdots$ & $-297377^{+148144}_{-111734}$ & $-1026290,\ -119690$ & $\cdots$ \\
$q\,(M_{\mathrm{comp}}/M_{\star})$ & $0.391^{+0.018}_{-0.018}$ & $0.356,\ 0.428$ & $\cdots$ & $0.490^{+0.024}_{-0.026}$ & $0.440,\ 0.539$ & $\cdots$ \\
\enddata
\end{deluxetable*}
\end{longrotatetable}

\begin{longrotatetable}
\begin{deluxetable*}{lcccccc}
\tablecaption{Astrometry Orbit Fits: HAT-P-14 AB and HAT-P-16 AB\label{tab:keck-hatp14-hatp16}}
\tabletypesize{\scriptsize}
\tablehead{
\colhead{} & \multicolumn{3}{c}{HAT-P-14 AB} & \multicolumn{3}{c}{HAT-P-16 AB} \\
\cline{2-4}\cline{5-7}
\colhead{} & \colhead{Posterior} & \colhead{95.4\% HDI} & \colhead{Prior} & \colhead{Posterior} & \colhead{95.4\% HDI} & \colhead{Prior}
}
\startdata
\multicolumn{7}{c}{Fitted Parameters}\\
\hline
$M_{\star}$ ($M_\odot$) & $1.380^{+0.031}_{-0.030}$ & $1.320,\ 1.441$ & $\mathcal{N}(1.380, 0.030)$ & $1.149^{+0.032}_{-0.029}$ & $1.089,\ 1.212$ & $\mathcal{N}(1.150, 0.030)$ \\
$M_{\mathrm{comp}}$ ($M_\odot$) & $0.190^{+0.010}_{-0.010}$ & $0.170,\ 0.211$ & $\mathcal{N}(0.190, 0.010)$ & $0.150^{+0.010}_{-0.010}$ & $0.130,\ 0.170$ & $\mathcal{N}(0.150, 0.010)$ \\
$a$ (AU) & $189^{+40}_{-57}$ & $125,\ 371$ & $1/a$ (log-flat) & $127^{+22}_{-42}$ & $82,\ 208$ & $1/a$ (log-flat) \\
$i$ ($^\circ$) & $102.0^{+6.5}_{-8.7}$ & $90.4,\ 148.8$ & $\sin(i)$, $0<i<180$ & $118.0^{+8.5}_{-11.3}$ & $103.9,\ 156.4$ & $\sin(i)$, $0<i<180$ \\
$\sqrt{e}\cos\omega$ & $0.764^{+0.226}_{-0.141}$ & $0.056,\ 0.997$ & $\mathcal{U}(-1, 1)$ & $0.325^{+0.594}_{-0.334}$ & $-0.791,\ 0.928$ & $\mathcal{U}(-1, 1)$ \\
$\sqrt{e}\sin\omega$ & $-0.321^{+0.265}_{-0.303}$ & $-0.774,\ 0.218$ & $\mathcal{U}(-1, 1)$ & $0.098^{+0.640}_{-0.409}$ & $-0.814,\ 0.809$ & $\mathcal{U}(-1, 1)$ \\
$\Omega$ ($^\circ$) & $87.756^{+6.199}_{-4.480}$ & $65.747,\ 100.844$ & $\mathcal{U}(0,360)$ & $353.281^{+55.988}_{-72.970}$ & $281.745,\ 360.000$\tablenotemark{d} &
$\mathcal{U}(0,360)$ \\
$\lambda_{\mathrm{ref}}$ ($^\circ$) & $274.068^{+16.185}_{-19.390}$ &
$233.748,\ 313.832$ & $\mathcal{U}(0,360)$ & $202.444^{+124.538}_{-39.541}$ & $0.000,\ 321.427$ & $\mathcal{U}(0,360)$ \\
Parallax (mas) & $4.447^{+0.010}_{-0.010}$ & $4.427,\ 4.468$ & $\mathcal{N}(4.447, 0.011)$ & $4.496^{+0.014}_{-0.015}$ & $4.466,\ 4.526$ & $\mathcal{N}(4.496, 0.016)$ \\
\hline
\multicolumn{7}{c}{Derived Parameters}\\
\hline
$P$ (yr) & $2078^{+670}_{-892}$ & $1070,\ 5631$ & $\cdots$ & $1254^{+335}_{-580}$ & $641,\ 2628$ & $\cdots$ \\
$e$ & $0.730^{+0.263}_{-0.158}$ & $0.271,\ 0.994$ & $\cdots$ & $0.552^{+0.328}_{-0.193}$ & $0.053,\ 0.899$ & $\cdots$ \\
$\omega$ ($^\circ$) & $336.335^{+32.628}_{-17.744}$ & $272.947,\ 360.000$ &
$\cdots$ & $13.707^{+80.977}_{-96.114}$ & $0.000,\ 179.877$ & $\cdots$ \\
$T_{\mathrm{p}}$ (BJD$-2450000$) & $-620522^{+278884}_{-215170}$ & $-1863954,\ -296759$ & $\cdots$ & $-221365^{+83868}_{-72034}$ & $-658798,\ -32163$ & $\cdots$ \\
$q\,(M_{\mathrm{comp}}/M_{\star})$ & $0.138^{+0.008}_{-0.008}$ & $0.122,\ 0.153$ & $\cdots$ & $0.130^{+0.010}_{-0.009}$ & $0.112,\ 0.150$ & $\cdots$ \\
\enddata
\tablenotetext{d}{For angular parameters reported over the range $0^\circ$--$360^\circ$, the $95.4\%$ HDI may cross the $0^\circ-360^\circ$ boundary because Orvara samples these angles over the extended interval $-\pi$ to $3\pi$.}
\end{deluxetable*}
\end{longrotatetable}

\clearpage
\onecolumngrid

\section{KPIC spectroscopy and relative RV}\label{sec:kpic_spectra}

\begin{figure}[h]
    \centering
    \includegraphics[width=0.9\textwidth]{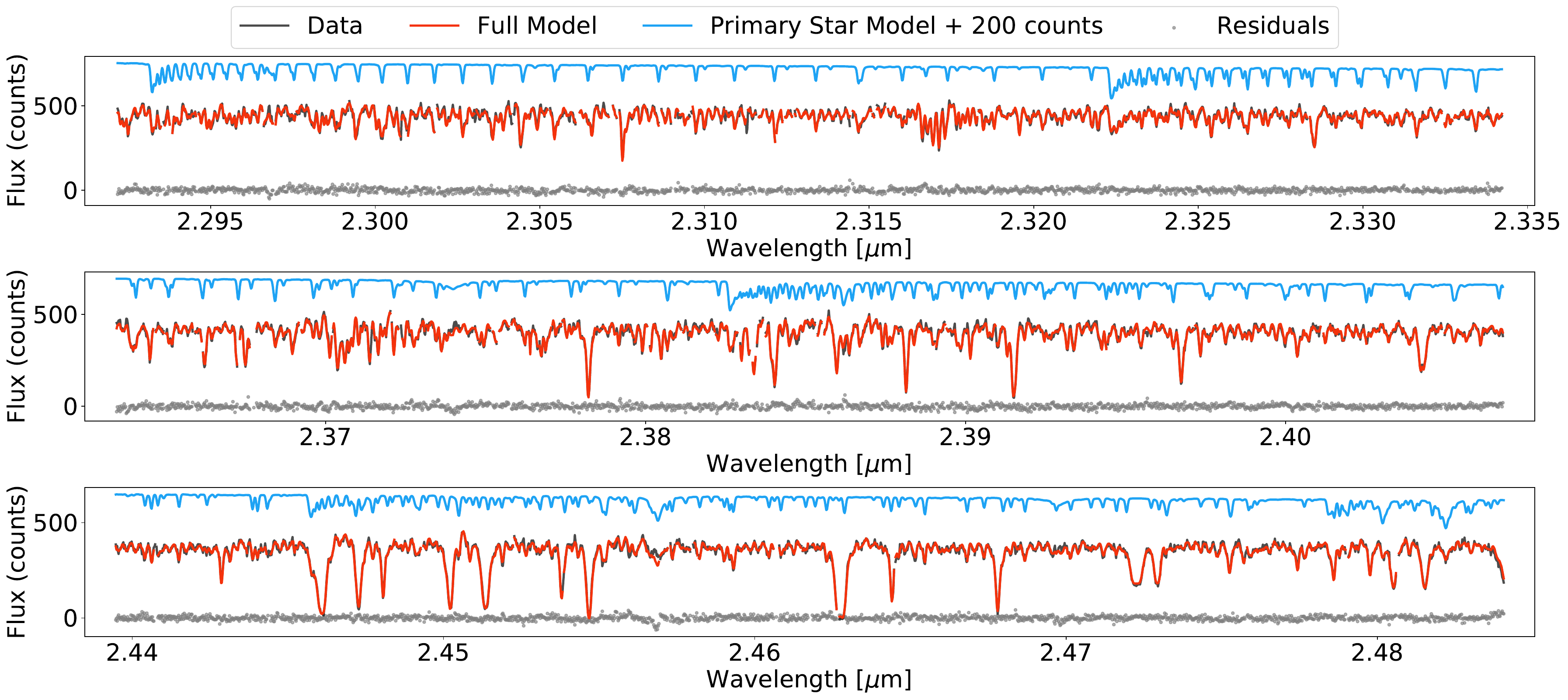}
    \vspace{0.3cm}

    \includegraphics[width=0.9\textwidth]{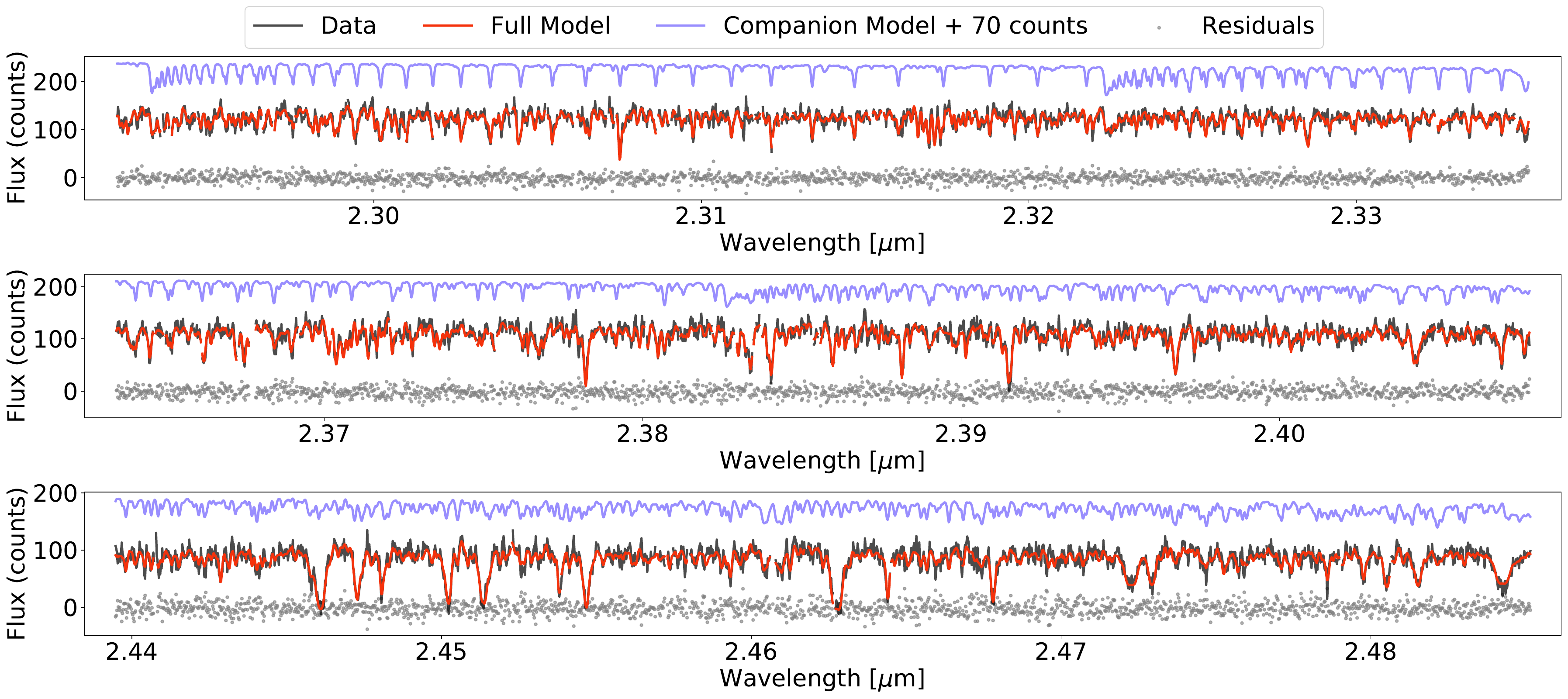}
    \caption{Top: KPIC spectrum (black) and BT-Settl model fit to WASP-2A,
    the primary star. The red model is the full model including the telluric
    transmission, whereas the blue model shown with an offset above is the
    star-only model. Bottom: Same but for WASP-2B, the secondary star. Here
    the purple shows the companion-only model.}
    \label{fig:wasp2_kpic}
\end{figure}

Below, we show the KPIC spectra and model fits to WASP-2A and WASP-2B in
Figure~\ref{fig:wasp2_kpic}. Besides the RV, we also measure
$\Teff=3615\pm50~{\rm K}$ and $\logg=5.05\pm0.25$ for the secondary, and
$\Teff=4950\pm50~{\rm K}$ and $\logg=5.10\pm0.25$ for the primary star.

\clearpage

\section{Radial Velocity Trends}\label{sec:rv_fit}

\begin{figure}[h!]
    \centering
    \includegraphics[width=0.95\textwidth]{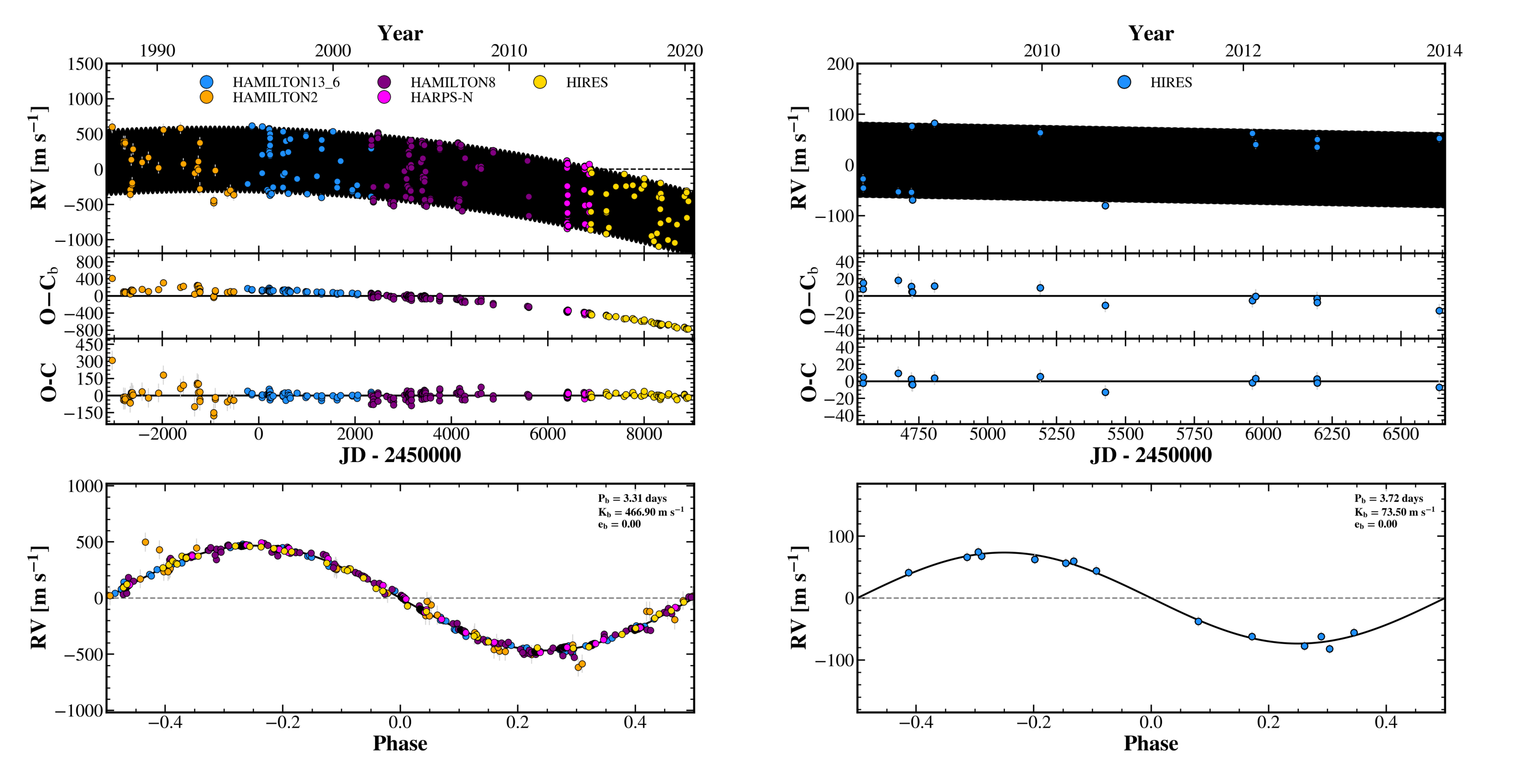}
    \caption{Best-fit models for $\tau$~Boo (left) and WASP-11 (right). A second order linear acceleration is preferred for $\tau$~Boo and a linear trend is preferred for WASP-11. The top panel shows the full RV time series. Below that are two sets of residuals: those after subtracting the inner hot Jupiter with the acceleration signal retained, and those after subtracting both the inner hot Jupiter and the Doppler trend. The phase-folded orbit of the inner hot Jupiter is displayed in the bottom panel.}
    \label{fig:Tau_Boo_HAT_P_10_RV_Fits}
\end{figure}

\noindent Here, we show the best-fit model fits to RV data of $\tau$~Boo and WASP-11. We find that a linear trend is favored for WASP-11 and curvature is favored for $\tau$~Boo. The RV residuals show clear trends that are consistent with the close-orbiting binary companions. We incorporate the residual RVs from Figure~\ref{fig:Tau_Boo_HAT_P_10_RV_Fits}, with the inner hot Jupiter signal removed and the acceleration signal retained, into the full orbit fits of the outer binary companion (see Section \ref{sec:HIRES_RV_Trends} and Figure~\ref{fig:Astrometric_Orbit_Fits}).

\clearpage

\newpage

\section{Additional Orbit Fit Parameters}\label{sec:Additional_Statistical_Tests} 

\begin{figure}[h!]
\begin{center}
{\includegraphics[width=0.95\linewidth]{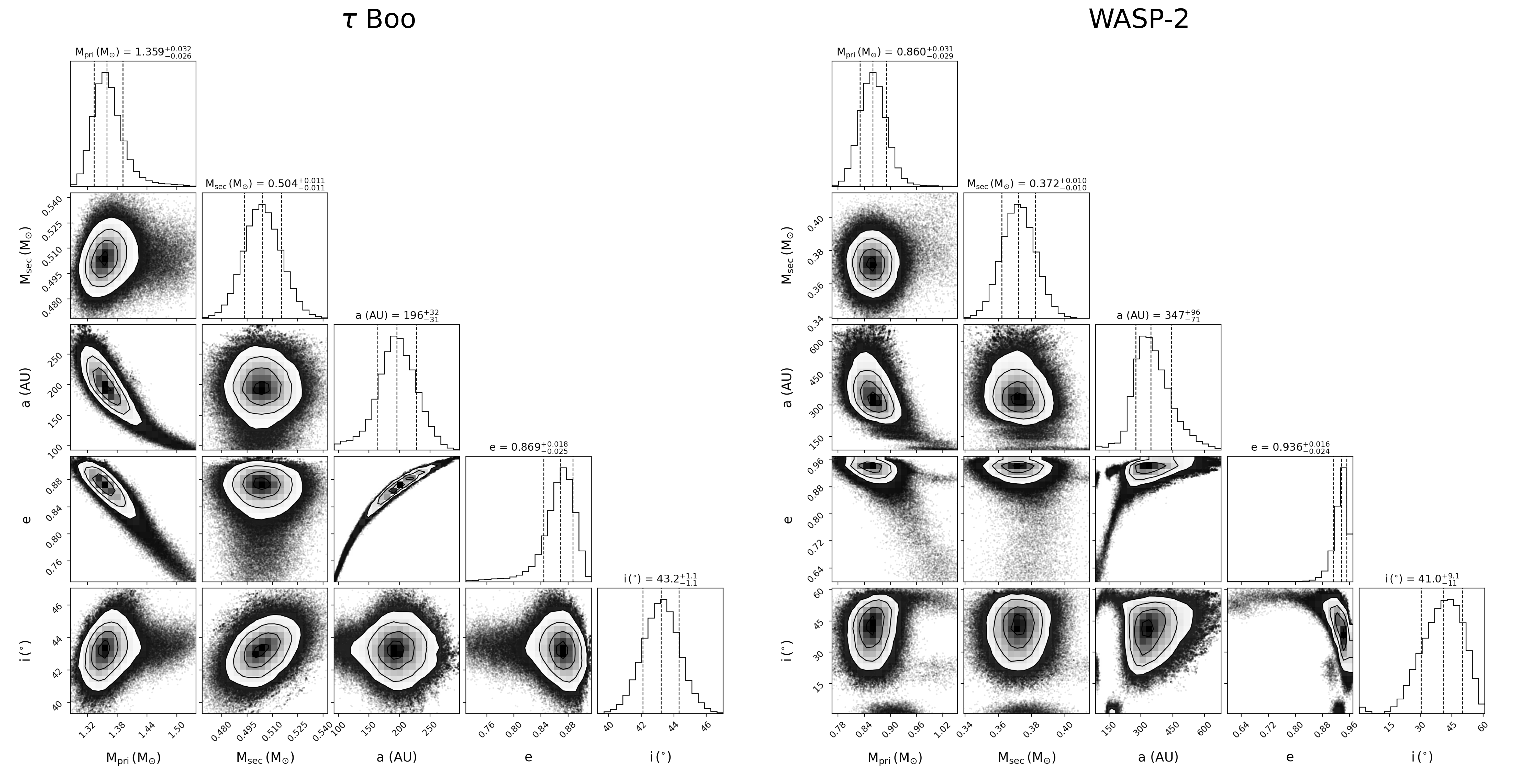}}
\caption{Joint posterior distributions and covariance between model parameters for the orbital fits for $\tau$ Boo and WASP-2. The derived parameters include mass of the primary star ($M_{\mathrm{pri}}$), mass of the secondary companion ($M_{\mathrm{sec}}$), semimajor axis ($a$), eccentricity ($e$), and inclination ($i$).}
\label{fig:Orvara_CORNER_TauBoo_WASP2}
\end{center}
\end{figure}

\begin{figure}[h!]
\begin{center}
{\includegraphics[width=0.95\linewidth]{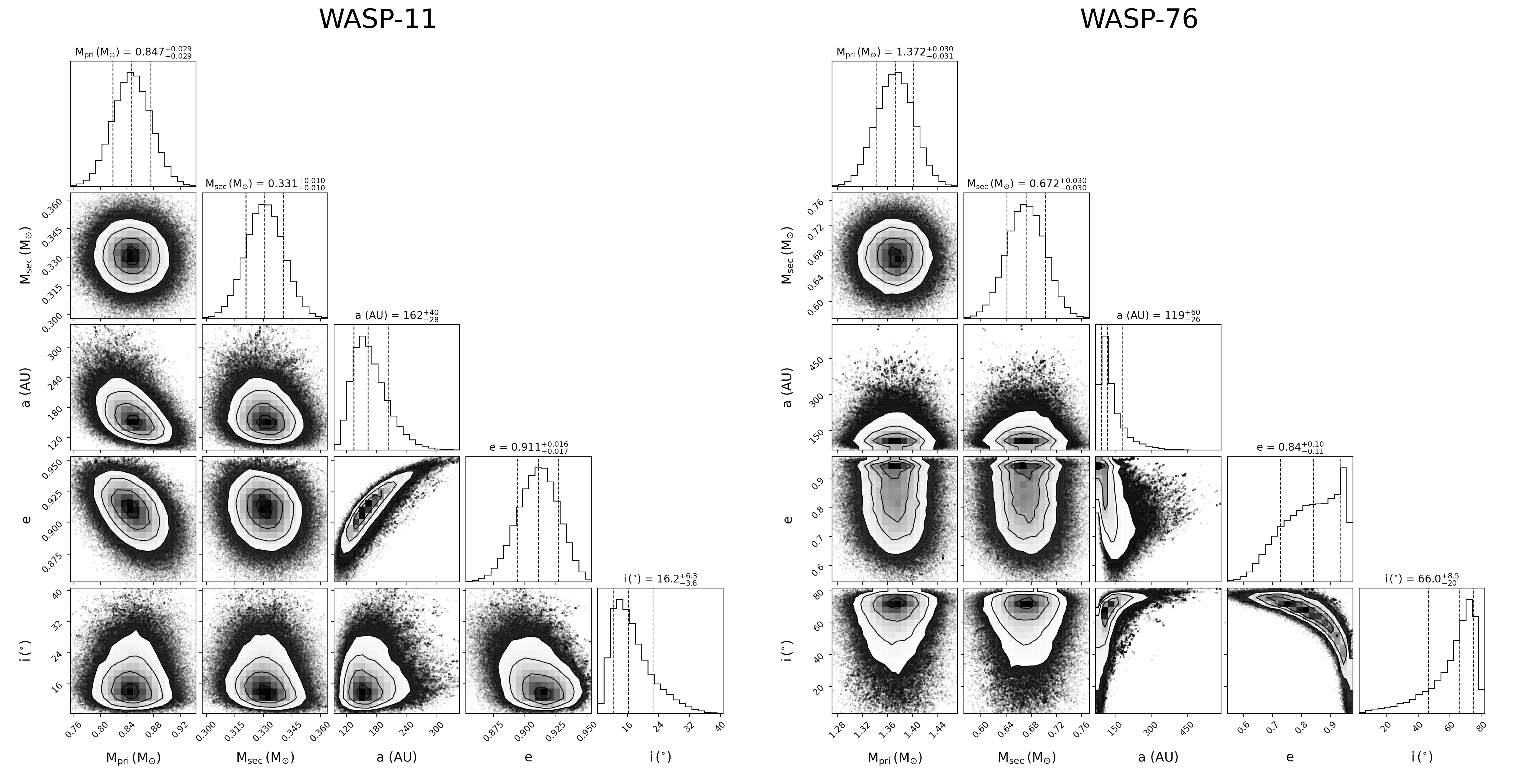}}
\caption{Joint posterior distributions and covariance between model parameters for the orbital fits for WASP-11 and WASP-76. See Figure \ref{fig:Orvara_CORNER_TauBoo_WASP2} for details.}
\label{fig:Orvara_CORNER_WASP11_WASP76}
\end{center}
\end{figure}

\begin{figure}[h!]
\begin{center}
{\includegraphics[width=0.95\linewidth]{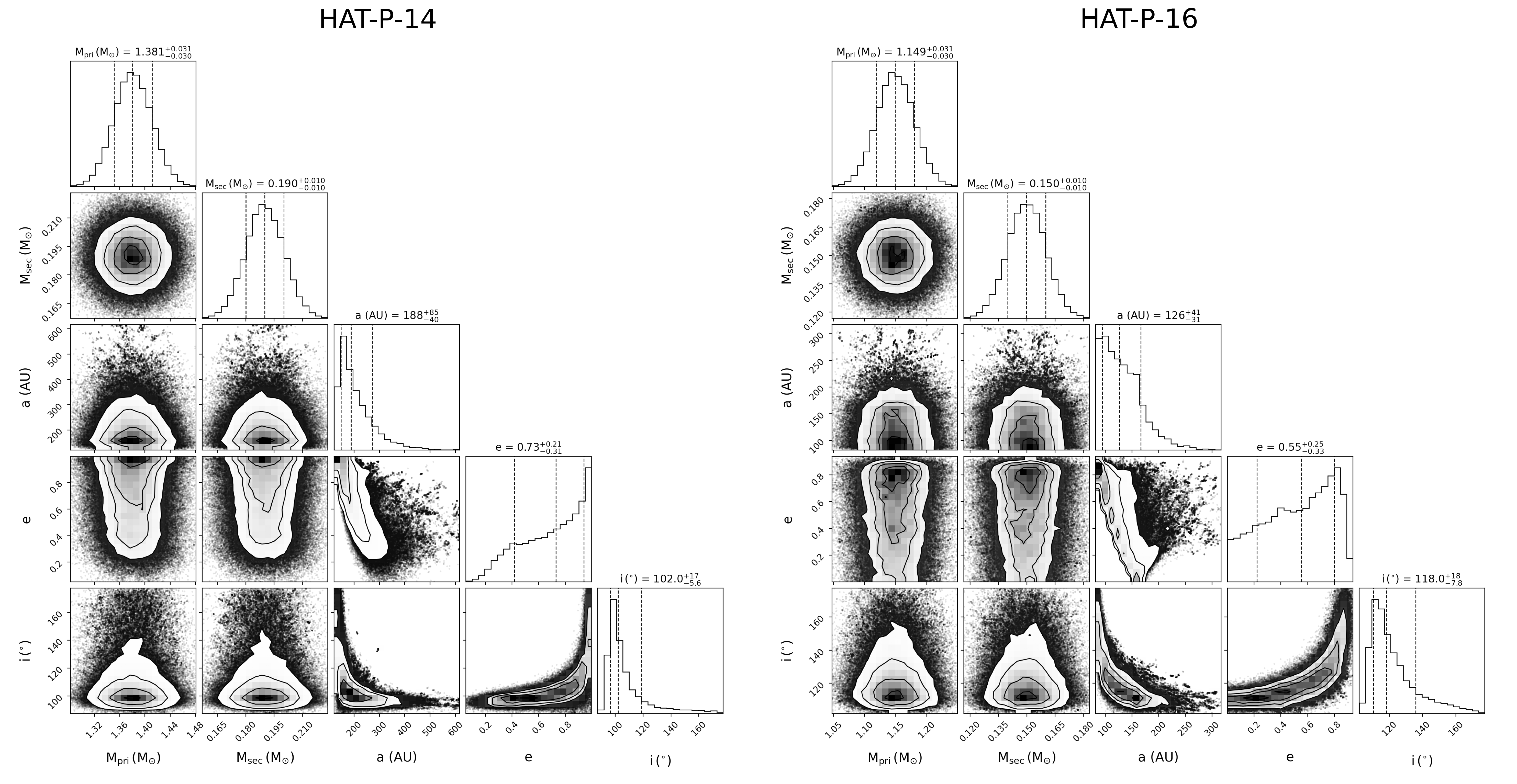}}
\caption{Joint posterior distributions and covariance between model parameters for the orbital fits for HAT-P-14 and HAT-P-16. See Figure \ref{fig:Orvara_CORNER_TauBoo_WASP2} for details.}
\label{fig:Orvara_CORNER_HATP14_HATP16}
\end{center}
\end{figure}

\begin{figure}[h!]
\begin{center}
{\includegraphics[width=0.95\linewidth]{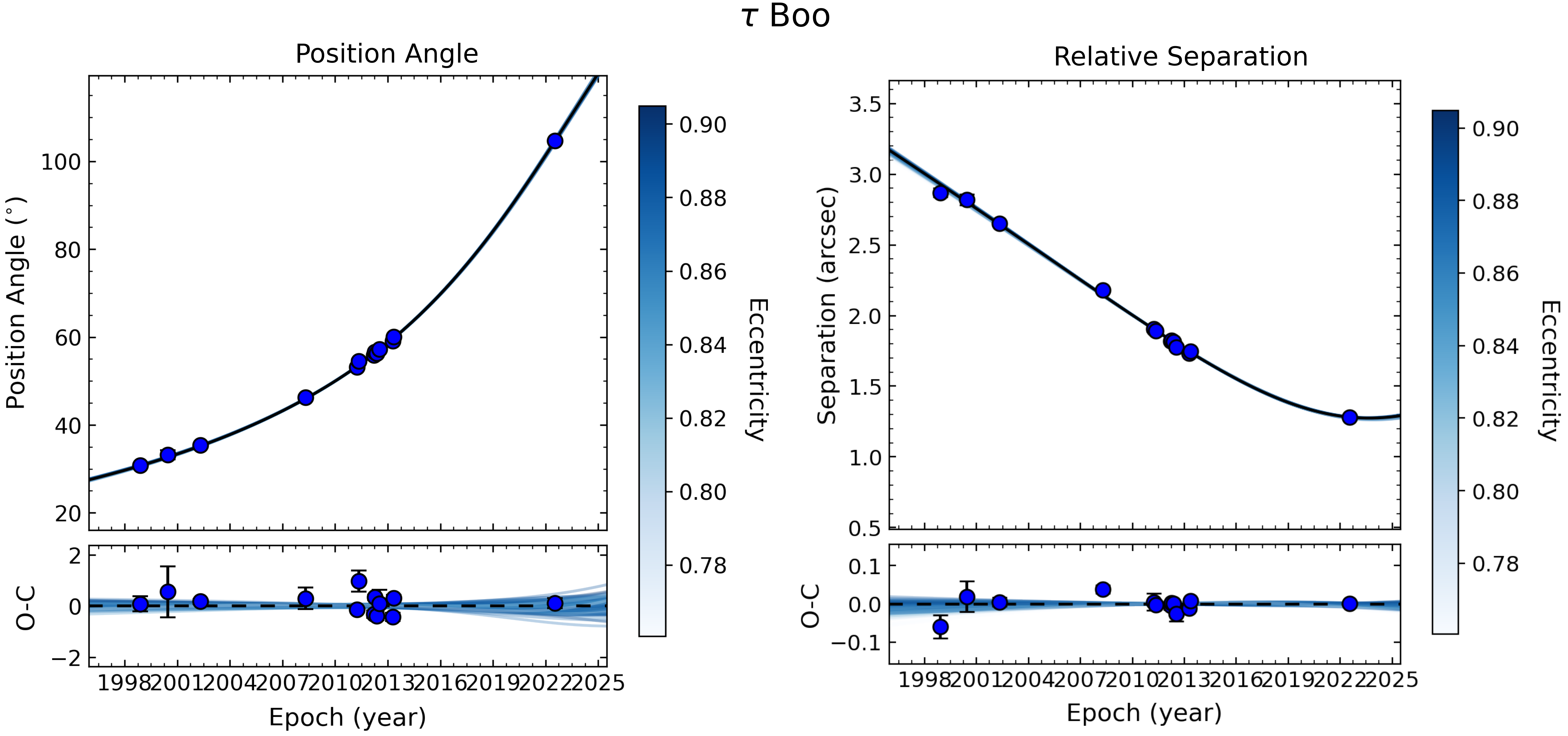}}
\caption{Left: position angle measurements and fits for $\tau$ Boo. Right: relative separation measurements and fits for $\tau$ Boo. The solid black line denotes the best-fit orbit from the MCMC chain are shown on top. The observed-calculated residuals are displayed below. 50 random orbit draws from the posterior orbit fits are colored by the eccentricity of the binary.}
\label{fig:Orvara_PA_RELSEPOC_TauBoo}
\end{center}
\end{figure}

\begin{figure}[h!]
\begin{center}
{\includegraphics[width=0.95\linewidth]{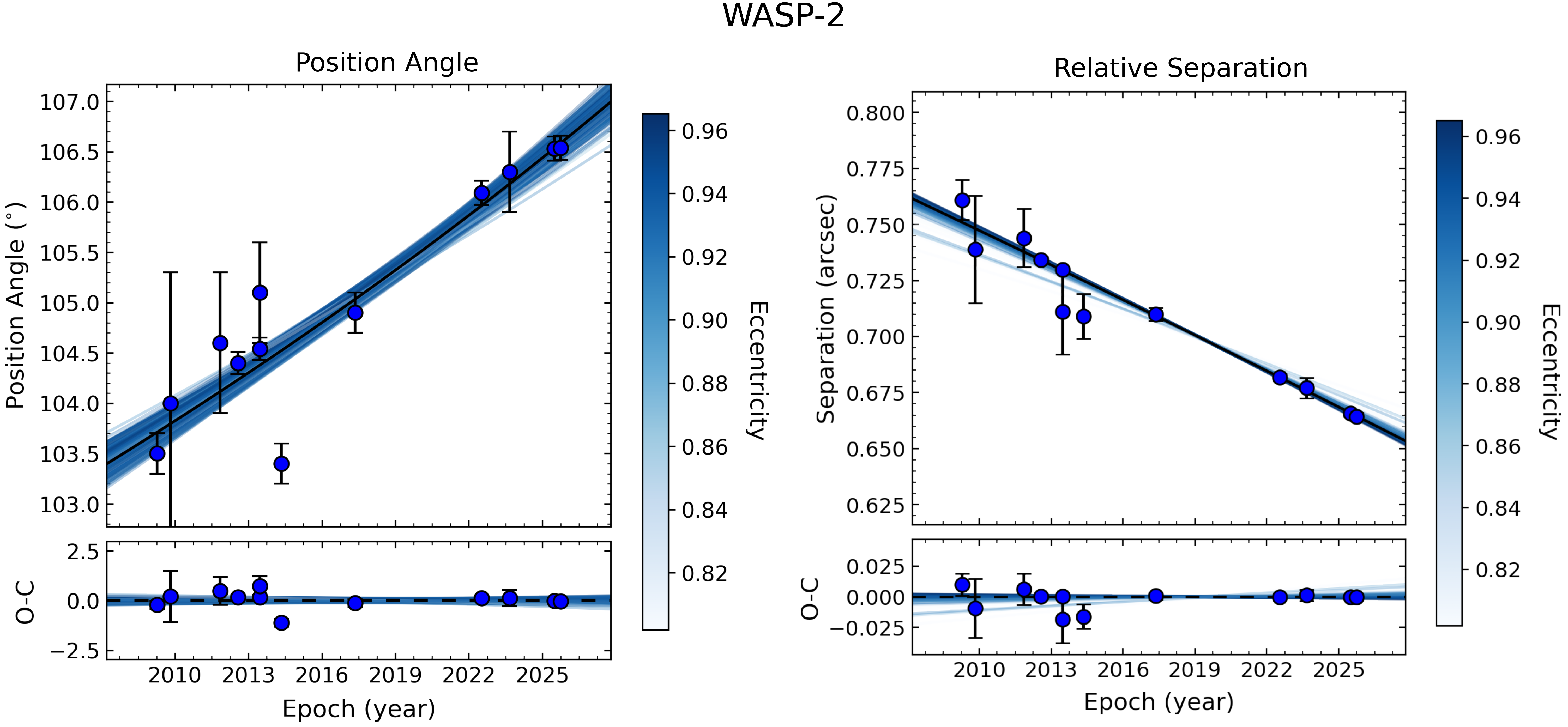}}
\caption{Position angle and relative separation measurements and fits for WASP-2. See Figure \ref{fig:Orvara_PA_RELSEPOC_TauBoo} for details.}
\label{fig:Orvara_PA_RELSEPOC_WASP2}
\end{center}
\end{figure}

\begin{figure}[h!]
\begin{center}
{\includegraphics[width=0.95\linewidth]{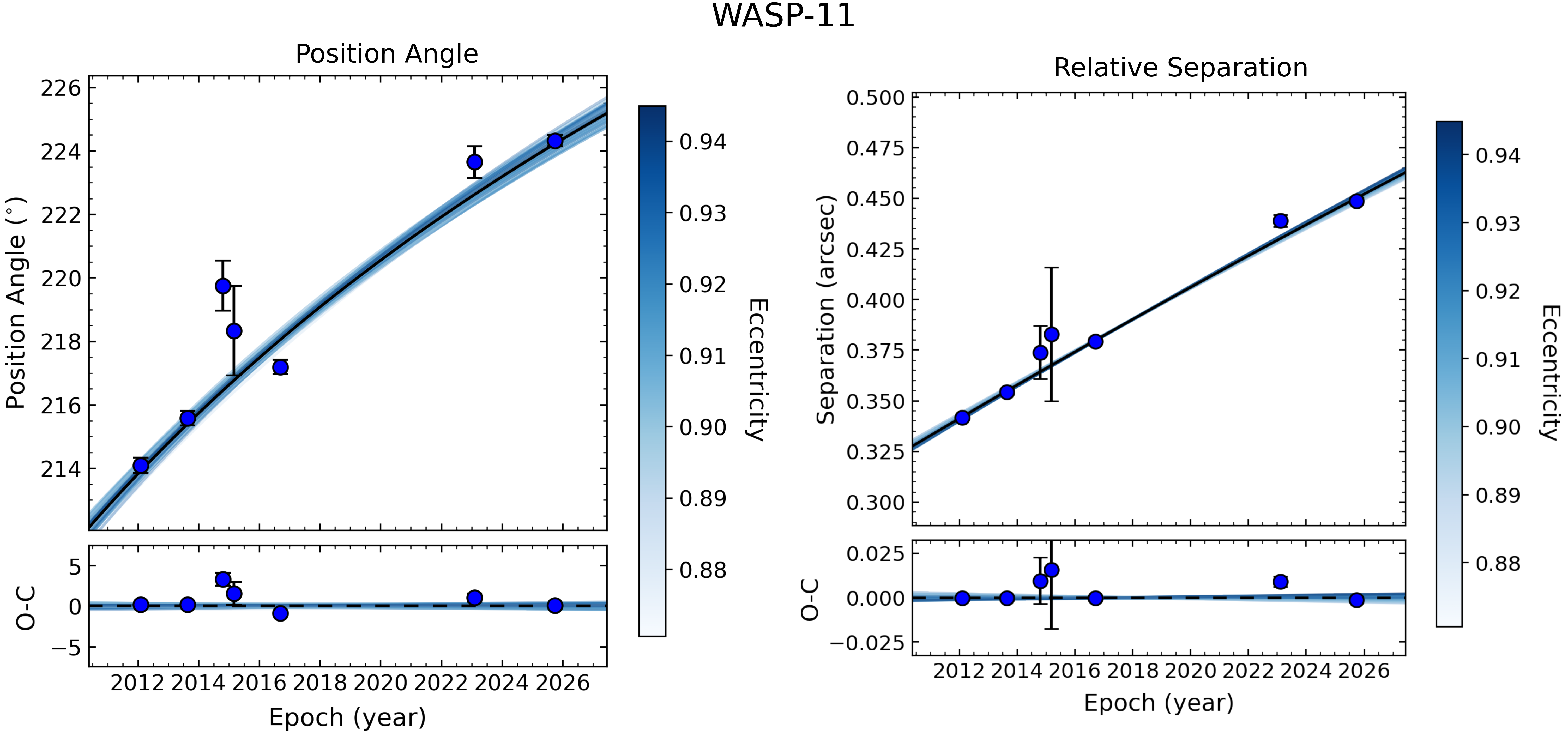}}
\caption{Position angle and relative separation measurements and fits for WASP-11. See Figure \ref{fig:Orvara_PA_RELSEPOC_TauBoo} for details.}
\label{fig:Orvara_PA_RELSEPOC_WASP11}
\end{center}
\end{figure}

\begin{figure}[h!]
\begin{center}
{\includegraphics[width=0.95\linewidth]{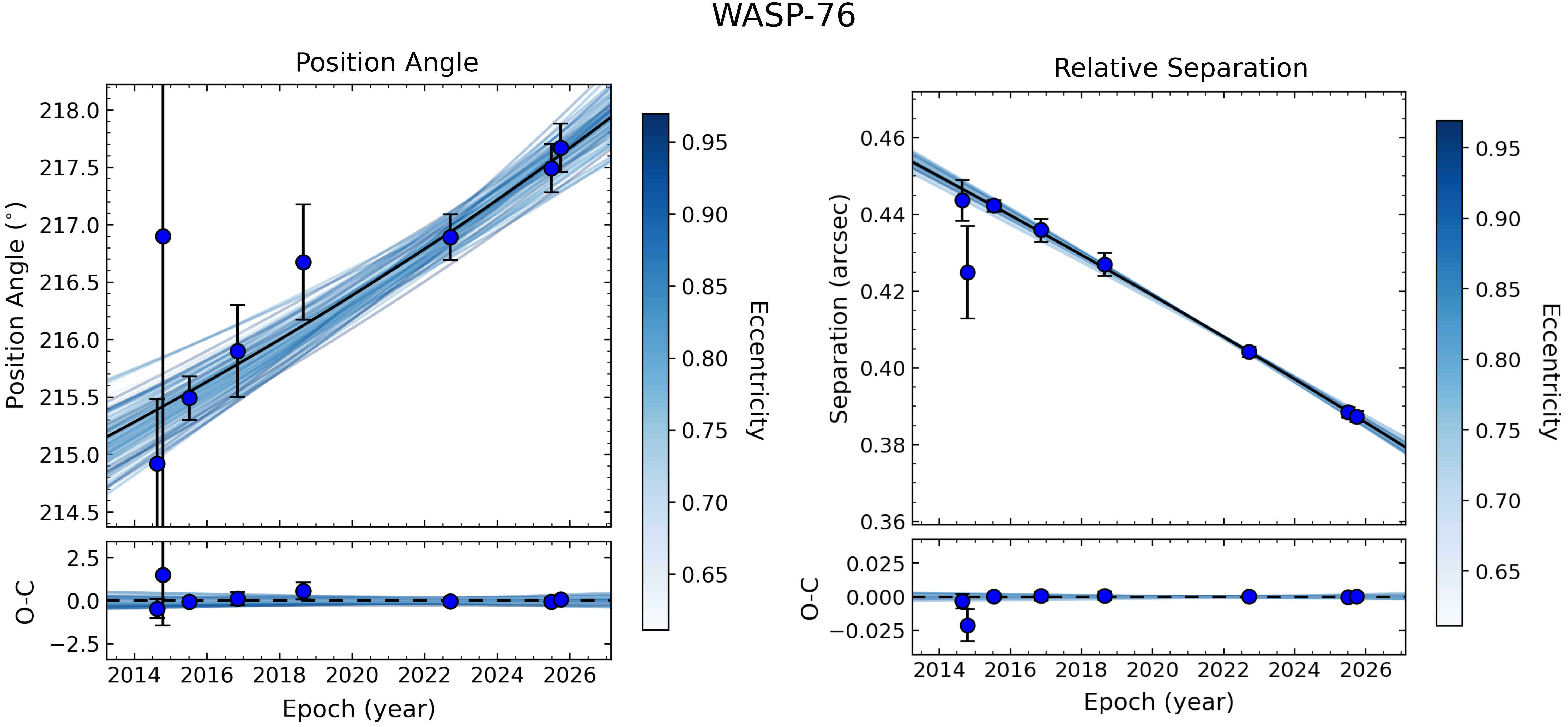}}
\caption{Position angle and relative separation measurements and fits for WASP-76. See Figure \ref{fig:Orvara_PA_RELSEPOC_TauBoo} for details.}
\label{fig:Orvara_PA_RELSEPOC_WASP76}
\end{center}
\end{figure}

\begin{figure}[h!]
\begin{center}
{\includegraphics[width=0.95\linewidth]{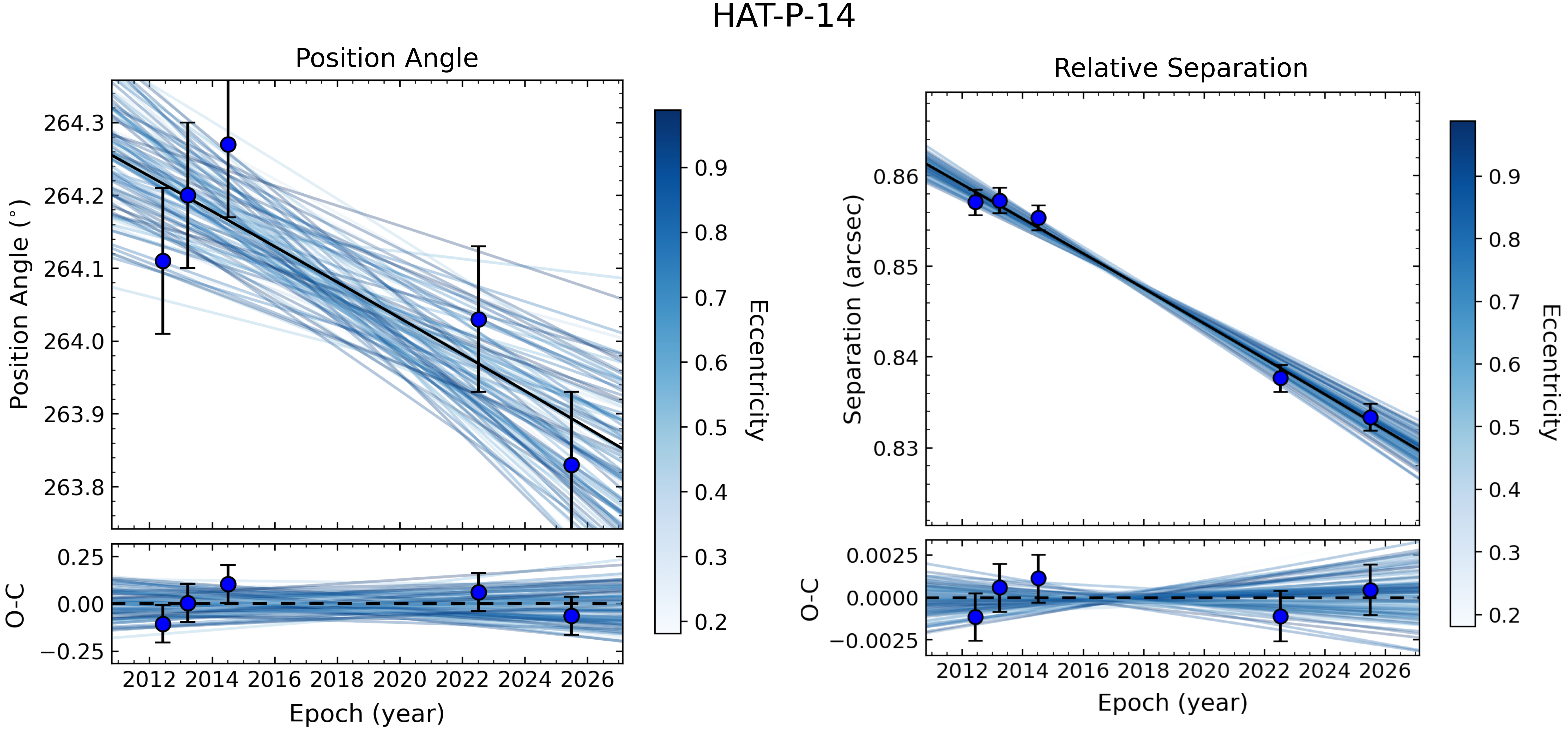}}
\caption{Position angle and relative separation measurements and fits for HAT-P-14. See Figure \ref{fig:Orvara_PA_RELSEPOC_TauBoo} for details.}
\label{fig:Orvara_PA_RELSEPOC_HATP14}
\end{center}
\end{figure}

\begin{figure}[h!]
\begin{center}
{\includegraphics[width=0.95\linewidth]{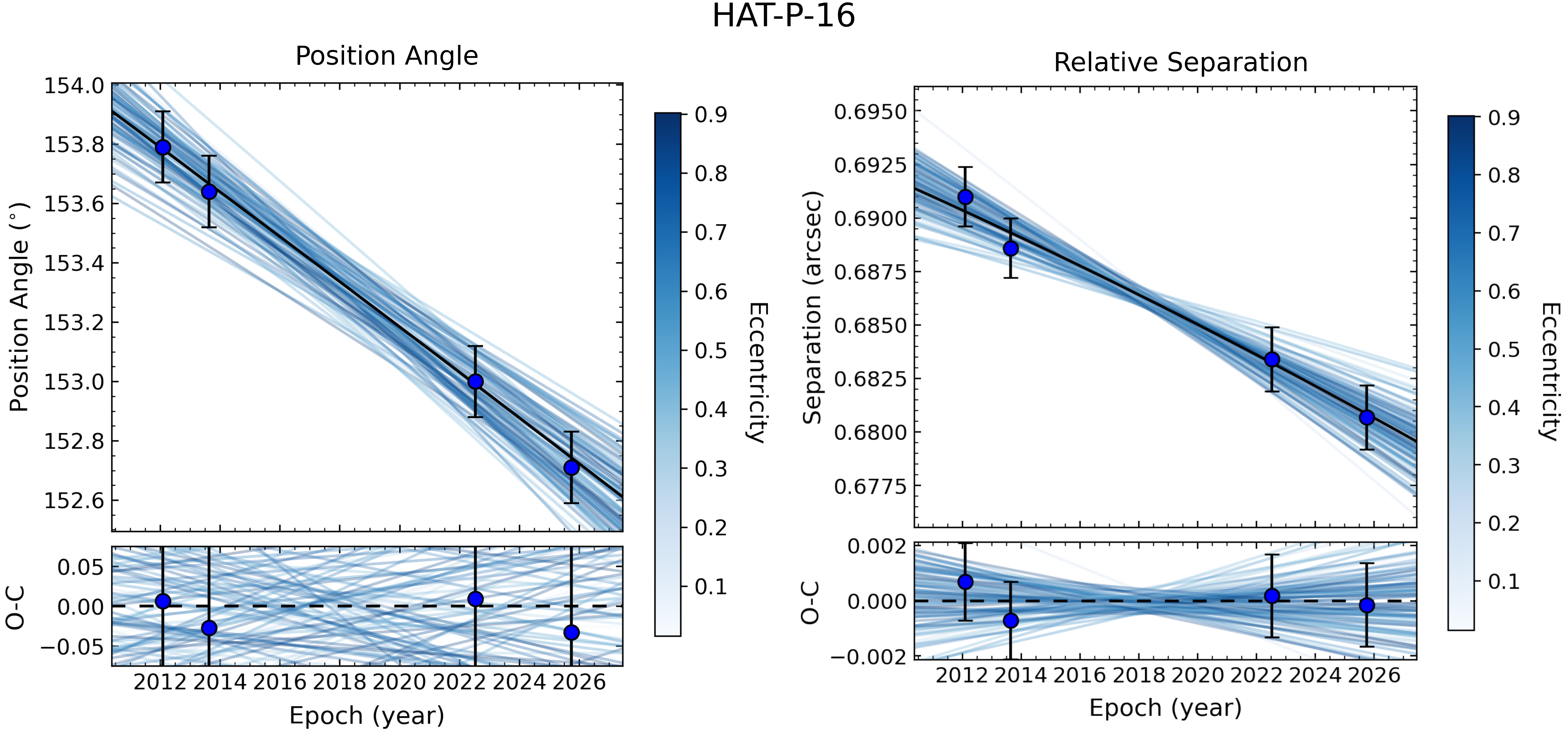}}
\caption{Position angle and relative separation measurements and fits for HAT-P-16. See Figure \ref{fig:Orvara_PA_RELSEPOC_TauBoo} for details.}
\label{fig:Orvara_PA_RELSEPOC_HATP16}
\end{center}
\end{figure}

\begin{figure}[h!]
\begin{center}
{\includegraphics[width=0.95\linewidth]{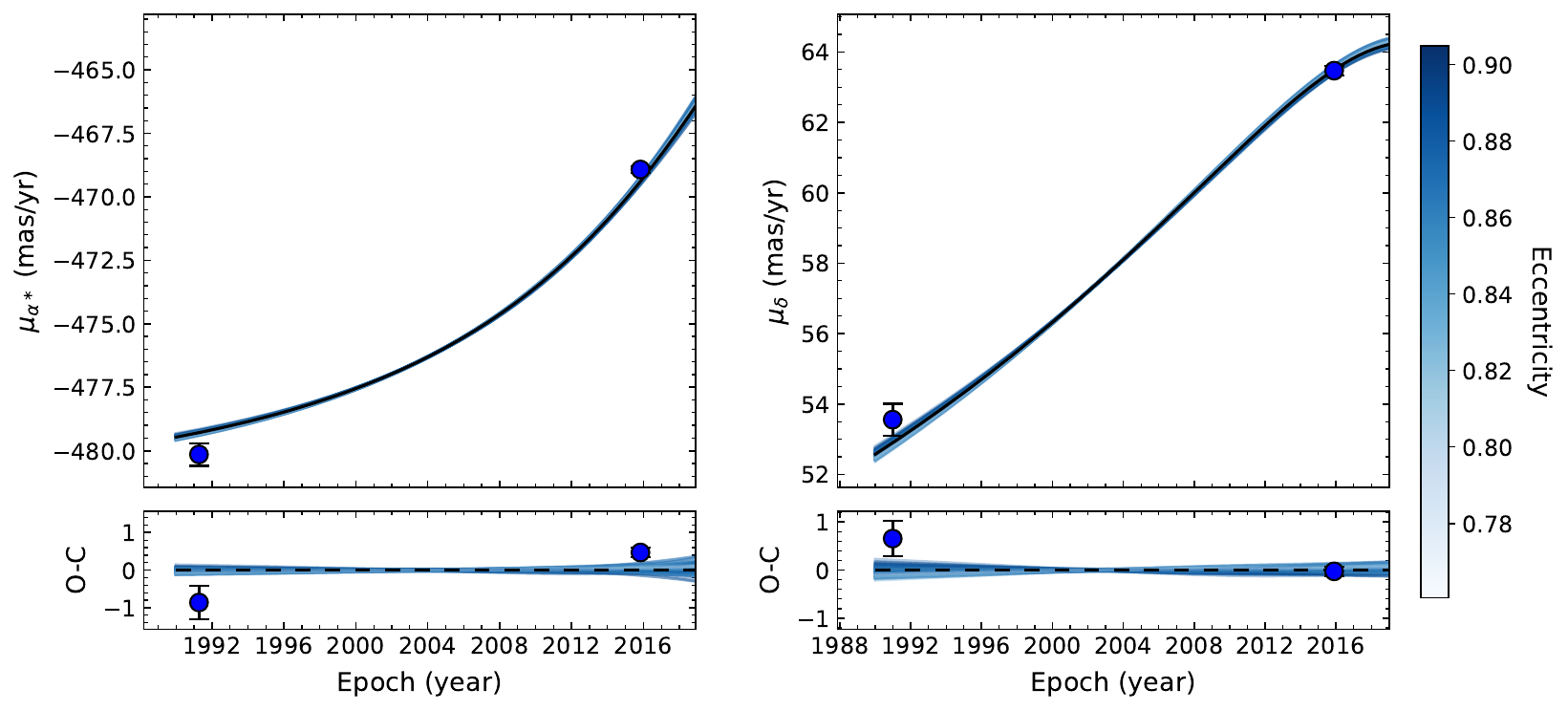}}
\caption{Top: Absolute astrometry measurements from \textit{Hipparcos}  and \textit{Gaia} and fits for the $\tau$ Boo system. Bottom: The observed-calculated residuals.}
\label{fig:Tau_Boo_HGCA}
\end{center}
\end{figure}

The joint posterior distributions between model parameters for the Keplerian fits for $\tau$~Boo AB, WASP-2 AB, WASP-11 AB, WASP-76 AB, HAT-P-14 AB, and HAT-P-16 AB incorporating all available data used in Figure \ref{fig:Astrometric_Orbit_Fits} is presented in Figures \ref{fig:Orvara_CORNER_TauBoo_WASP2}--\ref{fig:Orvara_CORNER_HATP14_HATP16}. In Figures \ref{fig:Orvara_PA_RELSEPOC_TauBoo}--\ref{fig:Orvara_PA_RELSEPOC_HATP16} we show the best-fitting Keplerian model and family of posterior draws for the relative astrometry as a function of time. Figure \ref{fig:Tau_Boo_HGCA} displays the model fits compared to the Hipparcos-Gaia proper motions.


\clearpage
\newpage
\bibliography{sample701}{}
\bibliographystyle{aasjournalv7}



\end{document}
